\RequirePackage{fix-cm}
\documentclass[smallextended]{svjour3}       % onecolumn (second format)
\smartqed  % flush right qed marks, e.g. at end of proof
\usepackage{amsmath}
\usepackage{amsfonts}
\usepackage{amssymb}\usepackage{graphicx}
\usepackage[ruled,vlined]{algorithm2e}
\usepackage{amsmath}
\usepackage{arydshln}
\usepackage[round]{natbib}
\usepackage[draft]{changes}
\begin{document}

\title{Multi-attribute community detection in International Trade Network%\thanks{Grants or other notes
}


\author{Rosanna Grassi         \and
        Paolo Bartesaghi \and
        Stefano Benati  \and
        Gian Paolo Clemente %etc.
}


\institute{R. Grassi (\emph{Corresponding author}) \at
              Department of Statistics and Quantitative Methods\\
              University of Milano - Bicocca \\
              Milano, Italy \\
              Tel.: +39-02-64483136\\
              https://orcid.org/0000-0001-5292-1723\\
              \email{rosanna.grassi@unimib.it}           %  \\
           \and
           P. Bartesaghi \at
              Department of Statistics and Quantitative Methods\\
              University of Milano - Bicocca\\
              Milano, Italy \\ 
           \and
          S. Benati \at
           School of International Studies\\ University of Trento\\
           Trento, Italy\\
           \and
          G.P. Clemente \at
          Department of Mathematics for Economics, Financial and Actuarial Sciences\\ 
          Universit\`{a} Cattolica del Sacro Cuore\\
          Milano, Italy \\           
}

\date{Published: 14 July 2021}

\maketitle

\begin{abstract}
Understanding the structure of communities in a network has a great importance in the economic analysis. Communities are indeed characterized by specific properties, that are different from those of both the individual node and the whole network, and they can affect various processes on the network. In the International Trade Network, community detection aims to search sets of countries (or of trade sectors) which have a high intra-cluster connectivity and a low inter-cluster connectivity. In general, exchanges among countries occur according to preferential economic relationships ranging over different sectors. In this paper, we combine community detection with specific topological indicators, such as centrality measures. As a result, a new weighted network is constructed by the original one, in which weights are determined taking into account all the topological indicators in a multi-criteria approach. To solve the resulting Clique Partitioning Problem and find homogeneous group of nations, we use a new fast algorithm, based on quick descents to a local optimal solution. The analysis allows to cluster countries by interconnections, economic power and intensity of trade, giving an important overview on the international trade patterns.

\end{abstract}

\section{Introduction}
\label{intro}
In network theory, a specific way to detect vertices having a peculiar common feature is termed clustering or community detection. 
Formally, a cluster, or a community, is a subgraph whose similarity or internal connections are stronger than the ones with the rest of the graph (\cite{Fortunato2010}). In recent years there was a surge of interest on the community structure in economic networks (\cite{Hajdu2019}) and, specifically, in international trade (\cite{Barigozzi2011,Garlaschelli2004,Garlaschelli2005,Li2003,Piccardi2011,Serrano2003,Serrano2007}).
The classical approach consists in finding sets of countries which are densely connected, through preferential economic relationships.    
A typical representation of this phenomenon is through a directed and weighted network, where nodes are countries and weighted links represent the aggregate trade flows. This representation is named in the literature as the International Trade Network (ITN). \\
Under this perspective, it becomes important to map the input-output interrelations among the countries through an inspection of the communities, where two countries share the same community if they have a comparable intensity in the trade flows or if they have preferential trade flows.\\
International trade has been widely studied in the literature showing that main characteristics  have  changed  over  time,  with  an acceleration of modifications occurring in the last decades. In particular, over the years, the composition of trade flows changed making countries even more deeply interconnected. The geographical distribution of trade also varied, with an increasing role of the emerging countries, especially in Asia.\footnote{$\rm https://www.wto.org/english/res\_e/publications\_e/anrep10\_e.htm$}

To detect the network structure, a key function is played by the vertex centrality.  The idea of centrality is quite simple to grasp: a numerical score is assigned to each node of the network so that the higher the score, the more central the node in the network. The literature has highlighted the importance to be central in an economic network (see \cite{Varela2015,Blochl2011,Barbero2016}). In particular, centrality may be associated with countries that are the most important hub of the ITN, even though they are not leading import or export countries (\cite{Blochl2011,DeBenedictis2011}). There are different metrics describing centrality, 
but it has been shown that different measures (degree, coreness, etc.)  identify different influential nodes \citep{Ferraz}. For instance, a node could be central if it is directly connected with many other nodes, if it has an intermediary role in communication, and so on. Indeed, there is no consensus on an univocal definition of network centrality, because each measure considers only one specific concept (see, e.g., \cite{Newman2010}). But, resorting to only one of them is discarding a large amount of the whole information available.
Related to centrality, the clustering coefficient is also an important index to measure the interconnections within a community. This coefficient has been developed in all the cases of weighted, unweighted, directed and undirected networks (see \cite{WasFaust,Watts_1998,Barrat_2004,Onnela_2005,CleGra,Fagiolo_2007}). In particular, \cite{Rotundo2010} discusses the clustering coefficient in presence of already established communities for directed networks and \cite{Roy} presents a concept of clustering coefficient which also includes the presence of missing indirect links in the construction of triangles. The association between communities and clustering coefficients is quite natural. Triangles are the easiest geometric visualization of communities, providing a picture of non-exclusive interactions among different agents. The relevance of this coefficient has been investigated also in the context of ITN (see, e.g., \cite{DeBenedictis2011,deb,Fagiolo2010,cepeda}). \\
As stressed in \cite{Barigozzi2011}, detecting the community structure of the ITN and how it correlates with country-specific variables and geography (e.g., distances between countries) is crucial from an international-trade perspective. Indeed, finding communities in the ITN means identifying clusters of countries that carry tightly interrelated trade linkages among them, while being relatively less interconnected with countries outside the cluster. 

In this work, we provide a new methodology for clustering countries based on a multi-criteria assessment of several topological indicators of centrality. The method consists of two steps. In the first step, we rank countries in ITN, according to various centrality measures. In the second one, based on those rankings, we compute the similarities between countries and then we apply the clustering algorithm based on the Clique Partition model.   

More specifically, in the first step, and unlike classical methodologies, we consider all the most prominent centrality definitions proposed in the literature that are relevant to international trade. 
Rather than advocate the superiority of one of them, we aggregate this rich multi-criteria assessment by defining a proper measure of similarity/dissimilarity between nations using their ranking positions. Next, we group together countries that have common structural features in terms of those rankings. The main advantage of our proposal is that we do not focus on a single and specific indicator of centrality, nor we come out with a detailed countries ranking. Rather, we are able to identify groups of countries that have similar structural properties in the ITN.
A specific tool developed for our project is a new heuristic algorithm to find clusters, based on the Clique Partition model (\cite{Grotschel1989,Grotschel1990,deAmorim1992}). 
The Clique Partition model consists of partitioning the vertices of a graph into the smallest number of cliques. First, a measure of similarity/dissimilarity between units must be established. This measure can take both positive and negative values, respectively if two units are similar or dissimilar. Then units must be partitioned in subsets, in such a way to maximize the similarity between them.
This model has some advantages over the classical $k$-means or hierarchical models. First of all, the clique partition model does not require either that the number of clusters were fixed in advance, e.g. the parameter $k$, or that the user should arbitrarily analyse the chart of the hierarchical clusters. Rather, the number of clusters results by the optimization of an objective function. Moreover, outliers are not forced to be in a cluster, but they can form peculiar groups of a single element. Finally, the principle of the method is that clusters are composed of mutually homogeneous data, while the $k$-means models first try to establish cluster's centres and then groups are composed by units that are similar to centres. Conversely, the clique partitioning forms groups of similar units. Experimental comparison between the clique partition and other clustering methods can be found in \cite{Wang2008}.  
The paper is organized as follows. In Section \ref{literature}, we recall main literature related to network theory, analysis of ITN and main solution methods for clique partitioning problems. In Section \ref{sec:model}, we describe the methodological framework and the integer linear programming problem. In Section \ref{sec:CP}, we define the maximum clique partition problem as well as the algorithm applied for identifying the optimal solution. In Section \ref{sec:num}, a numerical application is developed by using the paradigmatic case of the ITN. Conclusions follow in Section \ref{sec:conc}.

\subsection{Novelty and advantages of the proposed methodology}
\label{nov}

	%The classical meaning of community refers to nodes that are clustered on the basis of the intensity of the connections: the weighted connections relations between nodes as dense as possible, while its elements are weakly connected with those not belonging to such a set 
	The classical meaning of community refers to the clustering of nodes on the basis of the intensity of the connections between them: the community structure maximizes the density or the intensity of the connections between nodes inside each cluster, while members of different clusters are as weakly connected as possible (\cite{Newman2004}, \cite{FortunatoHric2016}). %This idea is grounded on the fact that these connections already exist and one has only to discover them. 
	The efforts of the literature have focused on finding new methodologies to detect communities under specific conditions (i.e. large or overlapping data, node attributed graphs, multilayer networks, and so on). %Several methods have been developed to detect this existing mesoscale structure. 
	Some methods are algorithm-based, such as hierarchical clustering or edge removal (\cite{Clauset2008hierarchical}). Others are based on the optimization of specific criteria over all possible network partitions. In this context, it is well-known the optimization of a modularity function according to Newman’s definition (\cite{Newman2004}). \\
	We go one step beyond this idea, applying a graph partitioning methods, e.g. the clique partitioning, to the graph in which arcs are weighted by node silimilarities. For instance, in term of centrality, nodes can be grouped together if they have strategic importance in transmitting information, or if they have similar power or control in the network. Moreover, our method is not limited to grouping nodes based on a single characteristic, but it is able to consider simultaneously more than one feature. %It is worth noting that the intensity of the existing links can be recovered as one of the selected features in the multi-criteria setting. 
	This aggregation %which is based on multiple nodes' features, is particularly useful in applications of several contexts. Indeed, our proposal 
	is general enough to be applied to various frameworks.\\ 
	%Focusing on    
	%It is worth briefly stressing the novelties and the advantages of our methodology with respect to other approaches applied to the same network. 
	%the problem of the identification of communities in the ITN was addressed, among others, by \cite{Piccardi2012}, \cite{Barigozzi2011} and \cite{BCG2020}.
	We show an application to the ITN. In this context, the identification of communities of countries has been addressed, among others, by \cite{Piccardi2012}, \cite{Barigozzi2011} and \cite{BCG2020}.
	In \cite{Piccardi2012}, the authors %focusing on the application of the classical maximum modularity criterion, manage to identify a mesoscale structure in the ITN; they show, for instance, 
	apply the classical maximum modularity criterion showing that the recognition of %this 
	the mesoscale structure is increasingly difficult due to the growing complexity and globalization of the international economic interactions. 
	The correlation between the world partition in communities obtained by a modularity criterion and geographical distances has been investigated also in \cite{Barigozzi2011}. The authors, both at an aggregate level and at a number of  commodity-specific levels, compare the two maximum modularity partitions of the input-output network and of the weighted network of the geographical closenesses. They find a high similarity between aggregate trade and geography-based communities, greater than, for instance, communities determined by regional trade agreements. They conclude that geographically-related factors explain the patterns of global trade more than political determinants. %Finally, 
	In \cite{BCG2020}, the authors %suggest to 
	interpret the ITN as a metric space by using two different distance measures that overcome the limitations of the shortest-path distance. They highlight strong interconnections between countries and identify communities as clusters of close countries in terms of such distances, according to a varying threshold. \\
	Our approach is instead aimed at applying a modularity criterion not to the immediate network of economic exchanges between countries but to a network in which the connection between countries is represented by a measure of similarity in the role they play within the global framework. This similarity measure exploits indicators of different nature and, as a consequence, our results will be less dependent on immediate factors which can affect a stronger or weaker relationship between pairs of countries, such as geographical proximity, trade agreements, common language or traditional partnerships. 
	%The methodology in the present paper provides a different point of view and results here obtained may be used to shed light on different aspects of the hidden structure of the ITN.
	%As also stressed before, 
	Taking into account the relevance of countries in the network, the methodology proposed in this paper provides a different approach for identifying clusters. Indeed, results here obtained may be used to highlight different aspects of the hidden structure of the ITN with respect to traditional community detection approaches. In particular, we aim at merging in the same community countries that have an analogous role in the network. Indeed, as emphasized in the literature (see \cite{Cingolani2017}) to shed light on a country’s participation in global trade, it is therefore important to understand where the country is positioned in the network. Although, there is a growing literature concerned with measures for assessing countries' position, typically main results show that rankings of countries vary across measures of centrality that are used. Our proposal instead aims at detecting communities of countries with similar relevance by aggregating several indicators and taking into account  peculiarities and heterogeneity of different measures.

\section{Related literature}\label{literature}

In this section we briefly remind the main literature related to network theory and International Trade, as well as clique partitioning problems and the main solution methods.

Network theory has been traditionally used in sociology and political science in order to investigate international trade relations, being an effective tool in revealing the core-periphery structure of the countries or in studying the impact of the globalization on the international trade structure (\cite{Snyder1979,Smith1992,Kim2002}).
The topological and statistical properties of the international trades, also in a time perspective, have been deeply studied in several works (see for instance, \cite{Serrano2003,Garlaschelli2007,FagioloR2008}).
More recently, complex networks have also been used to investigate  economic and financial implications of the world trade. For instance, Kali and Reyes (\cite{Kali2007,Kali2010}) study the country's role in the ITN deducing  important implications in terms of economic growth and explaining the phenomenon of financial contagion. Both international trade and financial integration patterns are investigated by Fagiolo et al. (\cite{Schiavo2010}).
Another important issue is the identification of communities in the trade network. Barigozzi et al. \cite{Barigozzi2011} deeply study the topology of the international trade multi-network, aiming at discovering its community structure. In \cite{tzekina2008}, the authors analyse the evolution of communities (``islands''): from two large trading communities, centred on UK and US, to a fairly heterogeneous ``archipelago'' of trade, that seems to reflect a phenomenon of globalization.
Finally, dissimilarities between different layers of an international trade multiplex network have been studied in \cite{Zhang2017}. The authors characterize each layer as a commodity network in a specific time period.

The definition of communities can be naturally associated with a partition in clusters, and one of the most important model of community detection is the clique partition.
The presence of communities inside the network is revealed by the modularity index (see \cite{Newman2004,SANTIAGO2017844}), that corresponds to the objective function of a clique partition model. By maximizing the partition modularity, one can determine the community structure of the network (\cite{newman2004Fast,clauset2004finding,blondel2008fast,Danon2006,Aloise2010}). The clique partition model, as a combinatorial approach to cluster qualitative data, had a methodological development independent of the problem of community detection, as it has been introduced in \cite{Grotschel1989,Grotschel1990,deAmorim1992,PATTILLO20139} and its applications range in many different fields (see, for instance, \cite{BUTENKO2005}). It has been recognized that it is a NP-hard problem, implying that the exact solution cannot be computed in polynomial time, unless P=NP. In practice, exact methods can solve instances that do not exceed one hundred nodes (\cite{Mehrotra1998,Aloise2010}), so that the use of heuristic procedure is necessary in our applications (\cite{SANTIAGO2017844,CHELOUAH2000}).

\section{The model}\label{sec:model}

In this section, we describe our methodology for clustering countries on the basis of the similarity attributes. \\
A network is described by a graph $G=(V,E)$ where $V$ and $E$ are respectively the set of $n$ vertices and $m$ links (or edges). Two nodes are adjacent if there is a link $(i,j)$ connecting them. The degree  $d_i$ of a node $i$ is the number of links incident to it. If a weight $w_{ij}>0$ is associated with each link $(i,j)$, a weighted graph $G=(V,E,W)$ is obtained, being $W$ the set of weights. 
In general, both adjacency relationships between vertices of $G$ and weights on the links are described by a nonnegative, real $n$-square matrix $\textbf{W}$. In the unweighted case, matrix $\textbf{W}$ is simply the classical binary adjacency matrix $\textbf{A}$, of entries $a_{ij}$, where $a_{ij}=1$ if $(i,j) \in E$, 0 otherwise.
Since we consider network without loops, $a_{ii} = 0$ (or $w_{ii} = 0$).
The $(i,j)-$element of the $k-$power of \textbf{A} is the number of walks of length $k$ from $i$ to $j$. The Laplacian matrix is defined as $\mathbf{L}=\mathbf{D}-\mathbf{A}$, where $\mathbf{D}$ is the diagonal matrix having the vertex degrees on the diagonal entries. \\   
A network is directed if each link is directed, that is an arc $(i,j) \in E$ means that there is a link starting from $i$ and ending in $j$. The in-degree $d^{in}_i$  (out-degree $d^{out}_i$) of a node $i$ is the number of arcs pointing towards (starting from) $i$. The degree  $d^{tot}_i$ of a vertex is then the sum of the in and out-degree.
In the directed case, matrices $\textbf{A}$, for a binary network, and $\textbf{W}$, for a weighted network, are not symmetric. 

\subsection{Network attributes and rankings}
\label{ss:nar}

We are interested in specific characteristics of the nodes, such as their centrality or their level of interconnection within the network. Since the network is weighted and directed, we need appropriate measures that take into accounts both weights and directions. Thus, according to the four dimensions classification of centrality indices in \cite{Brandes}, we focus on four class of network indicators, each one computed using both incoming and outgoing links. These are in and out-strength, in and out-clustering, hub and authority and Laplacian centrality.  

The strength (in and out) is the natural extension to the weighted and directed case of the degree centrality. It counts both the number of ties and their intensity. Formally, for a node $i$, we have:

\begin{equation}\label{str_in}
s^{in}_i=(\textbf{A}^T\textbf{W})_{ii}=\textbf{W}^T_i\textbf{1}
\end{equation}

\begin{equation}\label{str_out}	
s^{out}_i=(\textbf{AW}^T)_{ii}=\textbf{W}_i\textbf{1}
\end{equation}
where $\textbf{W}_i$ corresponds to the $i-th$ row of the matrix $\textbf{W}$.

In particular, in our application, the in-strength $s_i^{in}$ measures the total trade flows incoming to the country $i$, that is the import. The out-strength  $s_i^{out}$ measures the total trade flows outgoing from the country $i$, that is the export.

Clustering coefficient measures the tendency of a node to be well interconnected with its neighbours. 
Local clustering coefficient of a node $i$ counts the number of observed weighted directed triangles connected to $i$, divided by all its potential unweighted directed triangles:

\begin{equation}\label{clust_w_dir}
c_i(\mathbf{\tilde{W}})=\frac{\frac{1}{2}[(\mathbf{\tilde{W}}^{\left[\frac{1}{3}\right]}+(\mathbf{\tilde{W}}^T)^{\left[\frac{1}{3}\right]}]^{3}_{ii}}{d^{tot}_i(d^{tot}_i-1)-2d^\leftrightarrow_i},
\end{equation}

where $\mathbf{\tilde{W}}=[\tilde{w}_{ij}]_{i,j\in V}$ is the normalized weighted matrix whose elements are defined as  $\tilde{w}_{ij}=\frac{w_{ij}}{max(w_{ij})}$ and  $d^{\leftrightarrow}_i=\sum_{j\neq{i}}a_{ij}a_{ji}$ is the degree of bilateral arcs between the node $i$ and its adjacent nodes. 

As pointed out in \cite{CleGra} and \cite{Fagiolo_2007}, we have four types of directed triangles to which $i$ could belong. They generate four types of clustering coefficients, that can be separately computed.

Formula (\ref{clust_w_dir}) includes all the four coefficients described in \cite{Fagiolo_2007}. 
Nevertheless, the country $i$ is part of the in-type and out-type triangles, highlighting the presence/role of the node $i$ in import/export between its neighbouring countries. Thus, in our analysis, in-clustering  and out-clustering coefficients seem more appropriate in capturing the role of the node $i$ in the exchanges between the closest countries, distinguishing between import and export:
\begin{equation}\label{clust_w_in}
c^{in}_i(\mathbf{\tilde{W}})=\frac{\frac{1}{2}(\mathbf{\tilde{W}}^T\mathbf{\tilde{W}}^{2})_{ii}}{d^{in}_i(d^{in}_i-1)},
\end{equation}

\begin{equation}\label{clust_w_out}
c^{out}_i(\mathbf{\tilde{W}})=\frac{\frac{1}{2}(\mathbf{\tilde{W}}^2\mathbf{\tilde{W}}^T)_{ii}}{d^{out}_i(d^{out}_i-1)}.
\end{equation}

In order to model the influence, or the prominence, of a country in a global scenario of trade flows, the eigenvector centrality is the most suitable measure. The generalization of this measure to directed networks allows to associate with a node two status: authority and hubness. The idea arises in the context of web page search to rank the importance of a page \cite{Kleinberg1999}. A web page is an authority if it is pointed by many other pages. Hubs are pages that link to many authoritative pages.
Formally, let $a_i$ and $h_i$ be the authority and hub scores respectively. Then, the following relations hold:

\begin{equation}\label{auth}
a_i=(\textbf{W}^T\textbf{h})_i
\end{equation}
and 
\begin{equation}\label{hubs}
h_i=(\textbf{W}\textbf{a})_i
\end{equation}

\noindent where the vectors $\textbf{a}$ and $\textbf{h}$ collect respectively authorities and hubs scores of all nodes. 

By formulas (\ref{auth}) and (\ref{hubs}), definitions of hubs and authorities are characterized by a mutually reinforcing relationship: essentially, a good hub is a page that points to many good authorities; a good authority is a page that is pointed to by many good hubs. The use of these measures is motivated by their interpretation: on one hand, authorities are central countries as they import in turn from central countries. On the other hand, hubs are central as they export towards central countries. To compute the scores (\ref{auth}) and (\ref{hubs}), an iterative algorithm (HITS - Hyperlink Induced Topic Search) is proposed in \cite{Kleinberg1999}. Starting with initial score vectors $\textbf{a}^0$ and $\textbf{h}^0$, through the power iteration method on $\textbf{AA}^T$ and $\textbf{A}^T\textbf{A}$, the process converges to the principal eigenvectors $\textbf{a*}$ and $\textbf{h*}$ of the matrices $\textbf{AA}^T$ and $\textbf{A}^T\textbf{A}$.

The idea behind the Laplacian centrality is that the importance of a vertex $i$ is related to the network ability to adapt itself to the deletion of the vertex, i.e. its resilience. 
The Laplacian centrality of a vertex $i$ is reflected by the drop of the Laplacian energy of the network deriving by the deletion of $i$ from the network. According to \cite{Lazic2006}, the definition\footnote{It is noteworthy that an alternative definition of Laplacian energy has been provided in the literature (see \cite{Gutman2006}). Although this alternative definition has been widely explored in the literature, we focus on the original version defined in \cite{Lazic2006} because it is related to the Laplacian centrality measure.} of the Laplacian energy is: 

\begin{equation}\label{energy}
E_L(G)=\sum_k\lambda_k^2
\end{equation}

\noindent where $\lambda_k$ are the eigenvalues of the Laplacian $\mathbf{L}$. \\
Therefore, let $G_i$ the graph obtained by deleting the node $i$ from $G$, the Laplacian centrality is (see \cite{Qi2012}):

\begin{equation}\label{LaplCent}
l_i=\frac{E_L(G)-E_L(G_i)}{E_L(G)}=\frac{(\Delta E)_i}{E_L(G)}.
\end{equation}

\noindent where $E_L(G)$ and $E_L(G_i)$ are the Laplacian centralities computed on $G$ and $G_{i}$, respectively and $(\Delta E)_i$ measures the effect on the Laplacian energy of the network of the removal of $i$.
Since the denominator $E_L(G)$ has the same value for all vertices, we focus on the numerator $(\Delta E)_i$, that is always nonnegative for the interlacing property of the eigenvalues of the Laplacian matrix (see \cite{Haemers}).
The Laplacian energy can be re-expressed in terms of strength\footnote{In case of unweighted graphs, formula (\ref{energy2}) gives the result provided in \cite{Lazic2006}: $E_L(G)=\sum_k d_k\left(d_k+1 \right)=\sum_k d^{2}_{k}+2m$. The use of entries of the Laplacian matrix, instead of eigenvalues, is meaningful especially for large networks.} (see \cite{Qi2012}, Th. 1): 
\begin{equation}\label{energy2}
E_L(G)=\sum_ks_k^2+2\sum_ {k<j}w_{kj}^2.
\end{equation}

\noindent Hence, the difference  $(\Delta E)_i$ is: 
\begin{equation}\label{delta}
(\Delta E)_i=s_i^2+\sum_{k\in N(i)}(w_{ki}^2+2s_kw_{ki})
\end{equation}

\noindent where  $N(i)$ is the set of neighbours of the node $i$.
This expression allows the following interpretation of the Laplacian centrality of $i$. This centrality depends (in a quadratic way) on the strength and on the weights of the neighbours of $i$. As stressed in \cite{Qi2012} and \cite{Baruah}, compared with other standard centrality measures proposed for weighted networks (e.g. strength or betweenness centrality), the Laplacian centrality is an intermediate measure between global and local characterization of the importance of a vertex. 
The generalization to directed and weighted case follows\footnote{See \cite{Adiga} and \cite{Kissani} for two definitions of Laplacian energy for directed graphs.}, giving an expression for weighted and directed Laplacian centrality (in and out) $l_i^{in}$ and $l_i^{out}$ derived by formula (\ref{delta}).

In our analysis, we intend to aggregate different indicators. Indeed, as already stressed, each measure has peculiarities and characteristics that highlight various aspects of the exchange relations between countries. \\
This heterogeneity requires an approach that cannot be simply based on the direct comparison among extremely different measures. \\
Given that each index has specific unit measures and range of variations, we will focus on the various country centrality rankings rather than their  absolute values. More specifically, first we calculate the country rankings according to any index, then we cluster countries according to their positions on those rankings.
Indeed, each indicator induces a ranking which represents the structural importance of a single node in the network.
Rankings analysis allows us to compare more than one centrality simultaneously. The comparison will be developed by computing a distance function between rankings. In particular in this work we refer to the Minkowski distance, also known as $L_p$-norm distance.

Let us order the scores of each node obtained for each centrality measure $k$ and let $r_i^k$ be the position of the node $i$ with respect to $k$. The Minkowski distance $d({\bf r}_i,{\bf r}_j)$ is

\begin{equation}
d({\bf r}_i,{\bf r}_j)=||{{\bf r}_i}-{{\bf r}_j}||_{p}=\left( \sum_{k=1}^K\left| r_i^k-r_j^k \right|^p\right)^{1/p}  
\end{equation}

being ${\bf r}_i$ the rankings vector of node $i$, $K$ the number of considered centrality measures and $p$ any real value such that\footnote{Although $p$ can be any real number, when $p<1$ the formula does not define a metric, being the triangle inequality not satisfied.} $p \geq 1$.\\ 

This distance measure is commonly used in the literature for computing the dissimilarity of objects described by numeric attributes. It is a generalized distance metric that includes others as special cases. In fact, although theoretically infinite measures exist by varying the value of $p$, just three have gained importance (Manhattan distance for $p=1$,  Euclidean distance for $p=2$ and Chebyshev distance for $p\rightarrow \infty$). 

A remarkable feature of this distance consists in grouping more than one objects, namely it allows to consider all the network indicators simultaneously, producing a global fictitious distance between couple of nodes ranking.
Furthermore, this distance allows to exploit several values of $p$ in order to better highlight the general features of the analysed data (see \cite{DeAmorim2012,Rudin2009}). For instance,  \cite{Rudin2009} highlights how different configurations of data concentration can be caught varying $p$, so that Minkowski distance can be used for effectively tackling data analysis problems.

In our context, we use this distance to construct a complete network $K_n$ having the same node set and weighted adjacency matrix $\mathbf{\Omega}$, whose entries are defined as:

\begin{equation}\label{eq:omega}
\omega_{ij}= \left\{ 
\begin{array}{ll}
\frac{1}{1+d({\bf r}_i,{\bf r}_j)} & {\rm for}\ i\neq j \\
0 & {\rm for}\ i = j 
\end{array}
.\right.
\end{equation}
\noindent These weights range in $[0;1]$ and turn out to be effective in describing the similarities between countries. Indeed, the more two countries have a similar behaviour, the smaller is the distance and the higher is the weight. 

\subsection{The Maximum Clique Partition Problem}\label{sec:CP}

The Clique Partition (CP) problem, as applied to our model, is defined as follows.  The complete undirected graph $G = (V,E)$ is given, with $V = \{1,\ldots,n\}$. For each $(i,j)\in E$, gains/costs $g_{ij}$ are defined, which can take both positive and negative values. In our application, positive values of $g_{ij}$ are similarities, negative values are dissimilarities. Let $P = \{V_1, V_2, \ldots,V_q\}$ be a feasible partition of $V$ and let $\pi(V_k) = \sum_{i,j \in V_k} g_{ij}$ be the gains/costs sum of subset $V_k$, for $1 \le k \le q$. The CP problem consists of finding the node partition $P$ that maximizes the objective function $f(P) = \sum_{k = 1}^q \pi(V_k)$.

It is important to note that values $g_{ij}$ must be both positive and negative, otherwise there is no incentive to discard negative values and the best partition would be the total set $P = \{V\}$. Therefore, we calculate $g_{ij}$ as the difference between $\omega_{ij}$ (that are positive and bounded between 0 and 1) and benchmark values $\omega^*_{ij}$, representing a neutral threshold. Neutral thresholds are calculated as follows.  
Let $\omega = \sum_{ij} \omega_{ij}$ be the total network similarities and let $\omega_i = \sum_j \omega_{ij}$ the sum of similarities appointed to unit $i$. The probability that a unit $x$ of network similarity would be allocated to node $i$ is 
$Pr[\mbox{$x$ incident to $i$}] = \omega_i/\omega$. If similarity has no structure, that is, it is independent of pairs $(i,j)$ because data do not have clusters, then:

\begin{align*}
Pr[\mbox{$x$ incident to $i$} \cap \mbox{$x$ incident to $j$}] & \\ = Pr[\mbox{$x$ incident to $i$}]\times Pr[\mbox{$x$ incident to $j$}] & = 2 \omega_i \omega_j/\omega^2.
\end{align*}

Then, if similarities are independent, the expected similarity between $i$ and $j$ should be:
$\omega^*_{i,j} = \frac{\omega_i \omega_j}{\omega}$.
So, we can calculate gain/cost $g_{ij}$ as the difference between the actual and the  hypothetical similarity:
$g_{ij} = \omega_{ij} - \omega^*_{ij}$. In this way we obtain values $g_{ij}$ that are both positive and negative. 
The integer linear programming formulation of the Clique Partition is then:

\begin{equation}\label{cp:ilp}
\max\, \sum_{i \neq j}g_{ij} x_{ij}
\end{equation}

subject to

$$
\left\{ 
\begin{array}{c}
-x_{ij}+x_{ik}+x_{jk}\leq 1, \quad \forall i< j < k, \ \ i,j,k \in V \\ 
-x_{ik}+x_{jk}+x_{ij}\leq 1, \quad \forall i< j < k,   \ \ i,j,k \in V  \\
-x_{jk}+x_{ij}+x_{ik}\leq 1, \quad \forall i< j < k,  \ \  i,j,k \in V  \\
x_{ij} \in \{0,1\}, \quad i < j, \ \  i,j \in V 
\end{array}
\right.
$$

where $x_{ij}$ is equal to $1$ if two nodes are in the same cluster and $0$ otherwise.

We experimented very long computational times when we tried to solve it through Integer Linear Programming. Therefore, we implemented a heuristic procedure based on shrinking the vertices of the graph. Shrink is the subroutine by which we take two vertices, representing single units or clusters, and we merge them together to obtain a single cluster. Shrink is described in Algorithm \ref{algo:shrink}. Input is a data structure $G^h = <V^h, g^h, \pi^h>$, in which
$V^h$ is the active node set, each node representing a set of the partition, $g^h$ are the shrunken costs, defined for every pair $i,j \in V^h$, $\pi^h$ are the clique costs, defined for every active node $i \in V^h$.
Output is a data structure $G^q = <V^q, g^q, \pi^q>$ in which $|V^q| = |V^h| - 1$.
When we shrink $i,j \in V^q$, we delete $j$ from the active nodes, see Line 1, and the  clique profit $\pi_i^h$ of $i$ increases by the arc profit $g^h_{ij}$, while all others remain the same, see Lines 2 and 3. In the next steps, the profit of $i$ inherits the profits of $j$'s connections, see Lines 5-7. 

\begin{algorithm}[!h]
	\DontPrintSemicolon
	\KwIn{The data structure: $G^h = <V^h, g^h, \pi^h>$, the pair $i,j\in V^h$}
	\KwOut{The data structure: $G^q = <V^q, g^q, \pi^q>$}
	$V^q \gets V^h - j$\;
	$\pi^q \gets \pi^h$\;
	$\pi^q_i \gets \pi^h_i + g^h_{ij}$\; 
	$g^q \gets g^h$\;
	\For{$k \in V^h$}{
		$g^q_{jk} \gets 0$\;
		$g^q_{ik} \gets g^q_{ik} + g^h_{jk}$\;
	}
	\Return{$G^q$}\;
	\caption{{\sc Shrink}}
	\label{algo:shrink}
\end{algorithm}

Subroutine Shrink is used to join nodes or clusters every time we find an improvement of the objective function, that is, when we find a pair $(i,j)$ such that $g_{ij}^h > 0$. The procedure is described in Algorithm \ref{algo:greedy}.
At the beginning, Lines 1 and 2, the partition $V^q$ is composed of subsets of one element and the profits $\pi$ associated to them are null. Then, in the loop 3-9, the greatest profit $g_{ij}$ is selected and, if positive, vertices $(i,j)$ are shrunken. Otherwise, the algorithm stops. The objective function is calculated in Line 10.

\begin{algorithm}[!h]
	\DontPrintSemicolon
	\KwIn{The CP Problem, defined with input $V, g$.}
	\KwOut{The Clique Partition $V^q$, clique costs $\pi^q$, objective function $f^q$}
	$V^q \gets \{1, \ldots, n\}$\;
	$\pi^q \gets 0$\;
	\While{$stop =$ False}{ 	
		$g_{ij}^q \gets \max\{g_{kl}^q| k,l \in V^q\}$\;
		\If{$g_{ij}^q > 0$}{
			$G^h \gets \mbox{Shrink}(V^q, i, j)$\;
			$G^q \gets G^h$\;
		}
		\Else{
			$stop \gets \mbox{True}$\;
		}
	}  %while
	$f^q \gets \sum_{i\in V^q} \pi_i^q$\;
	\Return{$G^q, f^q$}\;
	\caption{{\sc Clique Partition}}
	\label{algo:greedy}
\end{algorithm}

We found that Algorithm \ref{algo:greedy} calculates quickly good quality solution. However, it can be the case that the selected partition is suboptimal. Therefore, we implemented a version of the Neighborhood Search procedure proposed in \cite{Brusco09}. The procedure starts with a feasible partition $P$, in our case the one calculated through Algorithm \ref{algo:greedy}. Then we select at random $k$ vertices of $V$ and try to relocate them to different clusters, searching for an improvement of the objective function. The procedure is repeated several time and for different values of $k$, until no improvement are found for many consecutive attempts. But in our data, we found that most of the times the results of Algorithm \ref{algo:greedy} were not improved.

\subsection{An overview of the Ranking Aggregation/Clique Partitioning procedure}

The next pseudo-code (see Algorithm \ref{algo:agpa}) summarizes the methodology that we are proposing:

\begin{algorithm}
	\DontPrintSemicolon
	Calculate rankings $\mathbf{r}^k$, for every centrality measure $k = 1, \ldots, K$ \;
	Calculate similarity/dissimilarity $\omega_{ij}$ between every countries pairs $i,j$.\;  
	Calculate the gain/cost $g_{ij}$ for all $i,j$ pairs.\;
	Solve the Clique Partition model whose input are $g_{ij}$'s. 
	\caption{{\sc Aggregation and Partition}}
	\label{algo:agpa}
\end{algorithm}

In Step 1, we have $K$ centrality measures, as defined in Subsection \ref{ss:nar}. For every measure $k$, ($k=1,...,K$), we obtain the ranking ${\mathbf r}^k$, whose element $r_i^k$ is the position of country $i$ in the ranking according to the measure $k$. In Step 2, we calculate values $\omega_{ij}$ according to Formula (\ref{eq:omega}). In Step 3, we calculate the gains/costs needed to define the Clique Partition model explained in Subsection \ref{sec:CP}. Lastly, in Step 4, we apply the Algorithm \ref{algo:greedy}.

\section{Numerical application}\label{sec:num}
\subsection{International Trade Network}

In this section, we apply the model previously described in order to study the structure of the ITN. We focus on a World Trade dataset, made available by the Observatory of Economic Complexity\footnote{See https://atlas.media.mit.edu/en/}. In particular, data regard the world trade database developed by the research and expertise centre on the world economy (CEPII) at a high level of product disaggregation. Original data are provided by the United Nations Statistical Division (UN Comtrade) and then the dataset is constructed by CEPII using an original procedure that reconciles the declarations of the exporter and the importer. This harmonization procedure enables to extend considerably the number of countries for which trade data are available, as compared to the original dataset (see \cite{CEPII}). In particular, we consider the last version published in 2017, based on  the Harmonized Commodity Description and Coding System,
and that provides aggregated bilateral values of exports for each couple of origin and destination countries. We focus on the aggregated data of the last available year, namely, 2014. 

\begin{figure}[ht]
	\centering
	\includegraphics[scale=0.4]{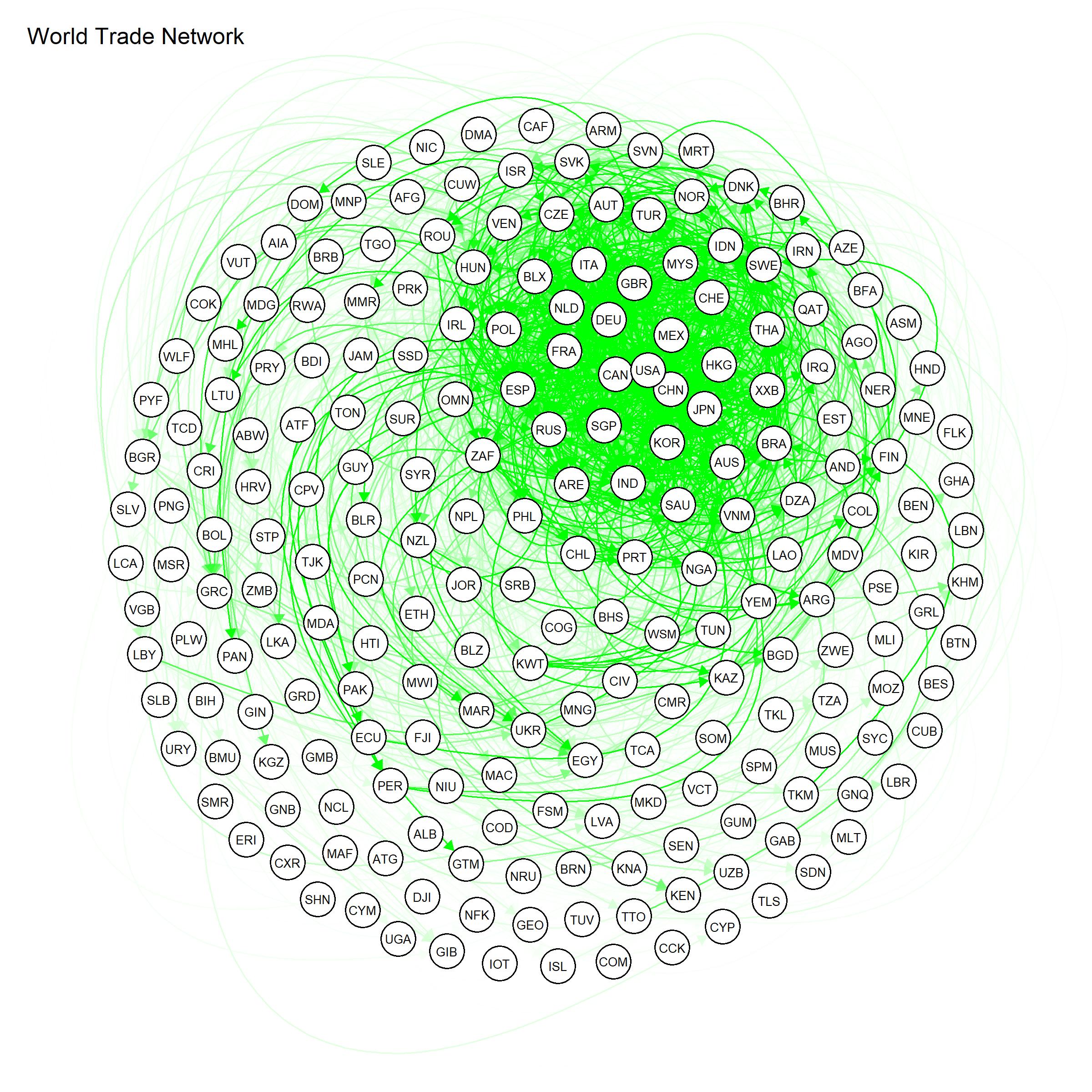}
	\caption{World Trade Network of imports and exports at the end of 2014. Each node is a country and weighted directed links represent the amount of product trades between couple of countries. Opacity of the link is proportional to the amount exchanged between countries.}
	\label{fig:network}
\end{figure}

\begin{figure}[ht]
	\centering
	\includegraphics[scale=0.5]{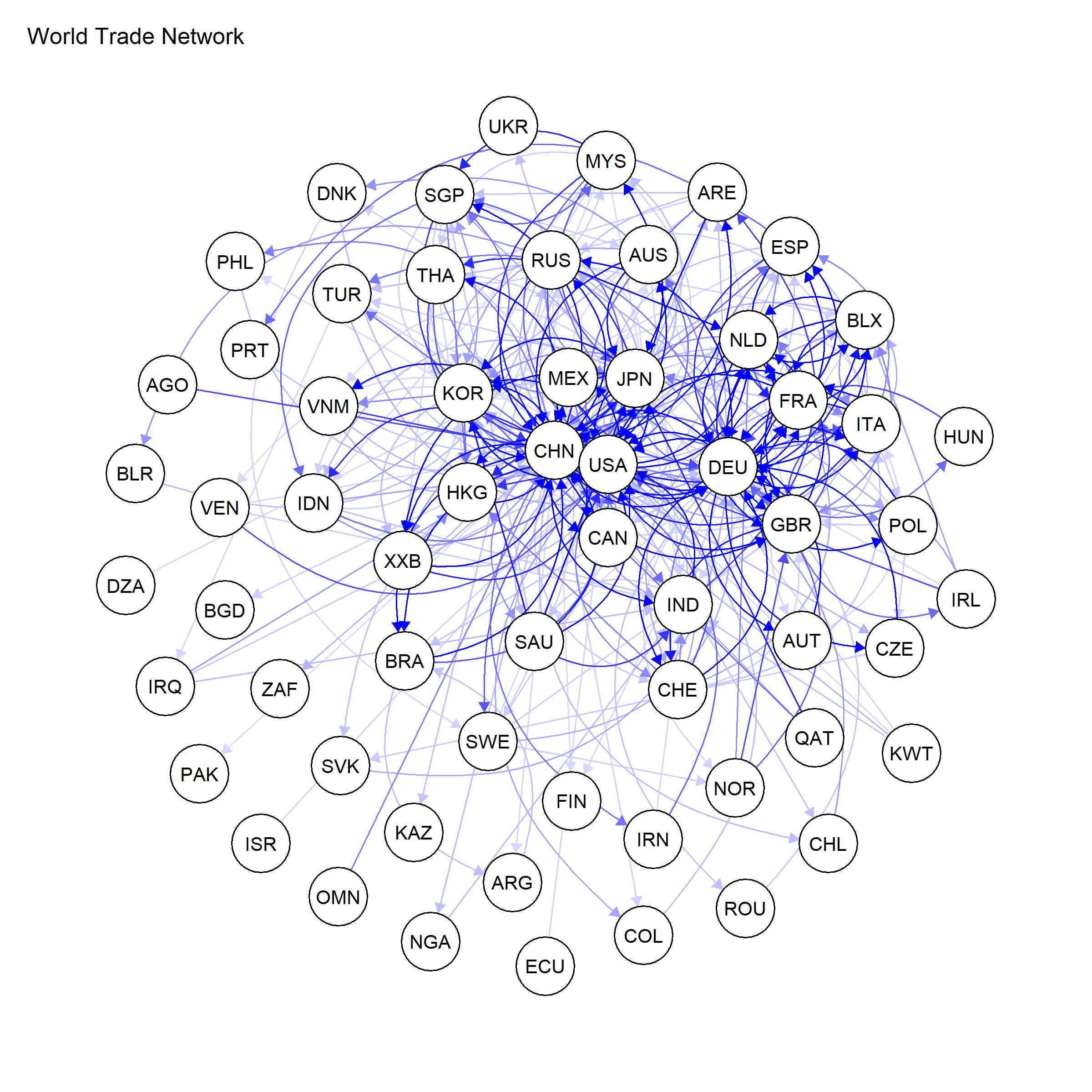}
	\caption{World Trade Network of imports and exports at the end of 2014. To highlight most relevant trades, we report only directed links whose weight is higher than 10 billion of US dollars. This amount approximatively corresponds to the quantile at level 99.3\% of the distribution of weights.}
	\label{fig:WTNzoom}
\end{figure}

\begin{figure}[ht]
	\centering
	\includegraphics[scale=0.5]{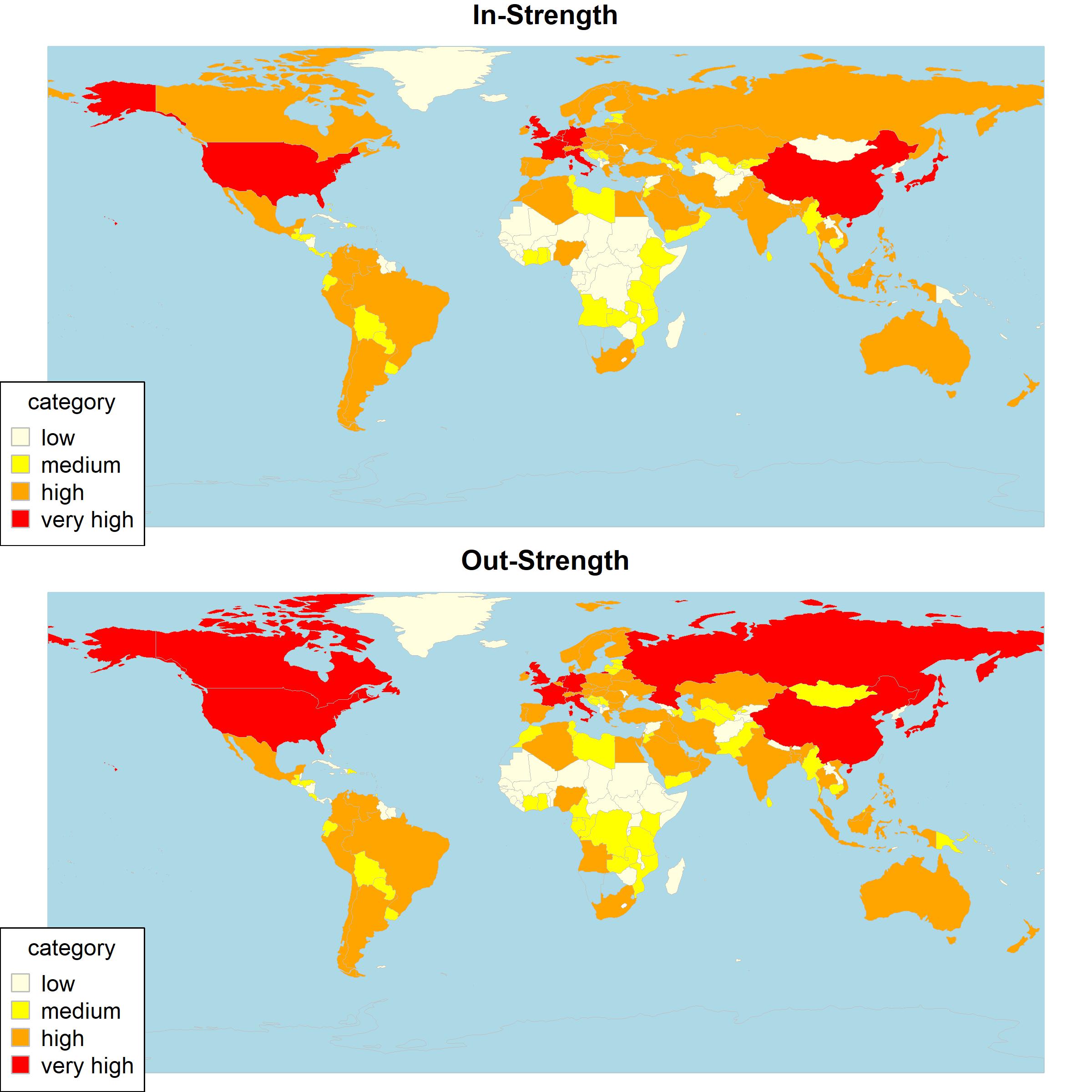}
	\caption{In and out-strength of countries in world trade network. Categories are based on the following classes $[0-q_{50}]$, $(q_{50}-q_{75}]$,$(q_{75}-q_{95}]$,$(q_{95}-q_{100}]$ where $q_{p}$ is the $p$-quantile of the in-strength  and out-strength distribution, respectively.}
	\label{fig:strength}
\end{figure}

Hence, we construct a directed and weighted network (see Figure  \ref{fig:network}), where each node is a country and weighted links represent the amount of product trades between couple of countries expressed in US dollars. This network is characterized by 220 countries and 26034 links. Its arc density is approximatively 0.54, because on average each country has a large number of trade partners and the entire system is intensely connected. However, the network is far from being complete or, in other words, most countries do not trade with all other countries, but they rather select their partners. Furthermore, world trade tends to be concentrated among a sub-group of countries and a small percentage of the total number of flows accounts for a disproportionally large  share  of world trade. We have indeed that, on average, each country has trades with more than an half of the other countries in the world, but the top 10 countries export more than 50\% of the total flow. To highlight most relevant trades, we report in Figure \ref{fig:WTNzoom} directed links whose weight is higher than 10 billion of US dollars.  The network is, in this case, characterized by 61 countries and 330 links. Additionally, key importers and exporters, classified in terms of strength, are displayed in Figure \ref{fig:strength}. Differences between import and export ranking are remarkable. United States, China, Japan, South Korea and some European countries (namely, France, Germany, Italy, Netherlands and United Kingdom) are world largest importers and exporters. Russia and Canada display instead a top ranking in terms of volume of exports. In particular, Russia is characterized by a significant positive trade balance, equal to approximatively 30\% of its total exportations. \\
Furthermore, as expected, greater  countries have more partners and they account  for  a  generally  larger  share  of world trade.  However, the relationship between the economic size and the number of partners is far from perfect, as indicated by the  correlation, around 0.5, between the total value of (in or out) flows and the number of partners for each country.

\subsection{Numerical results and discussion}

\begin{figure}
	\centering	
	\includegraphics[scale=0.3]{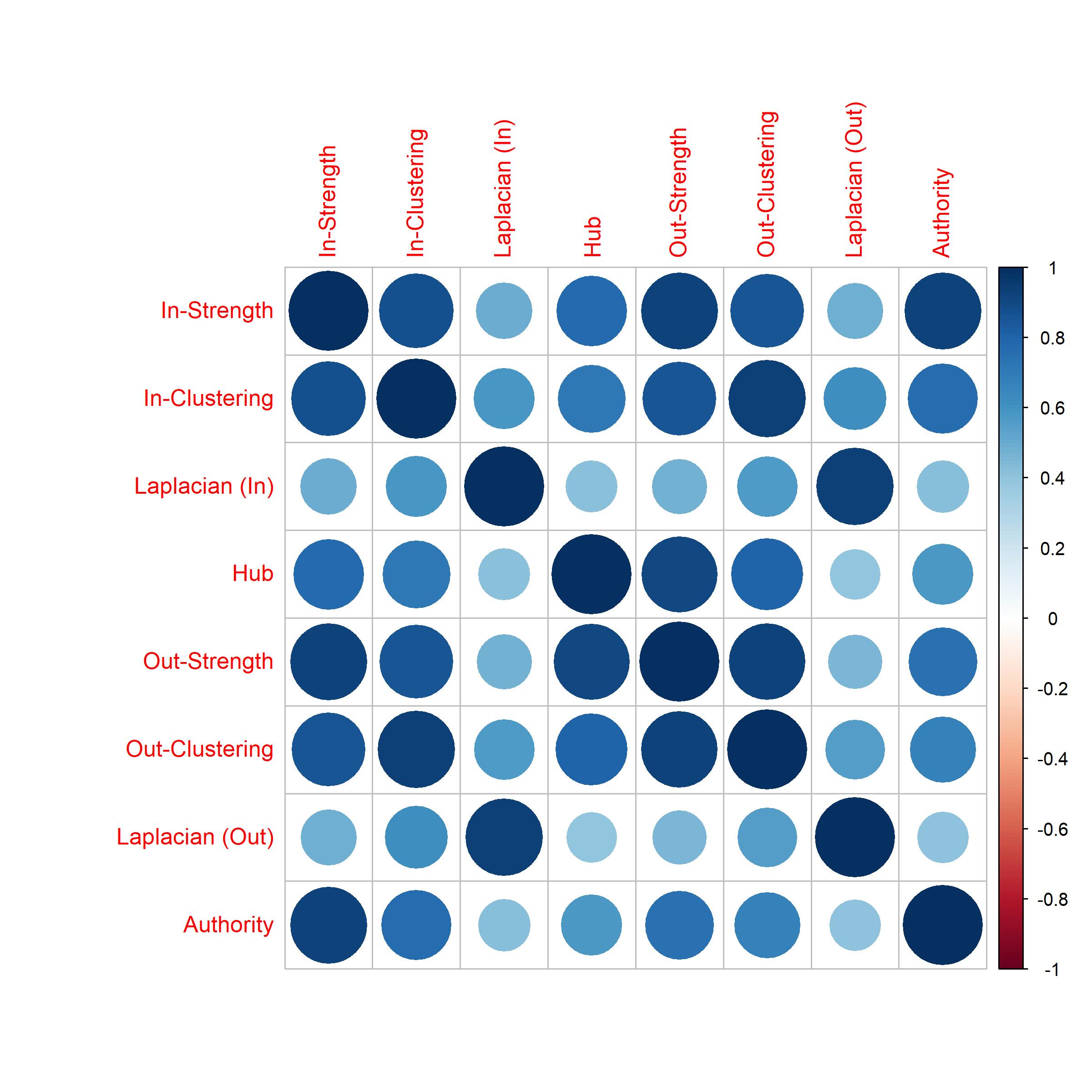}
	\includegraphics[scale=0.25]{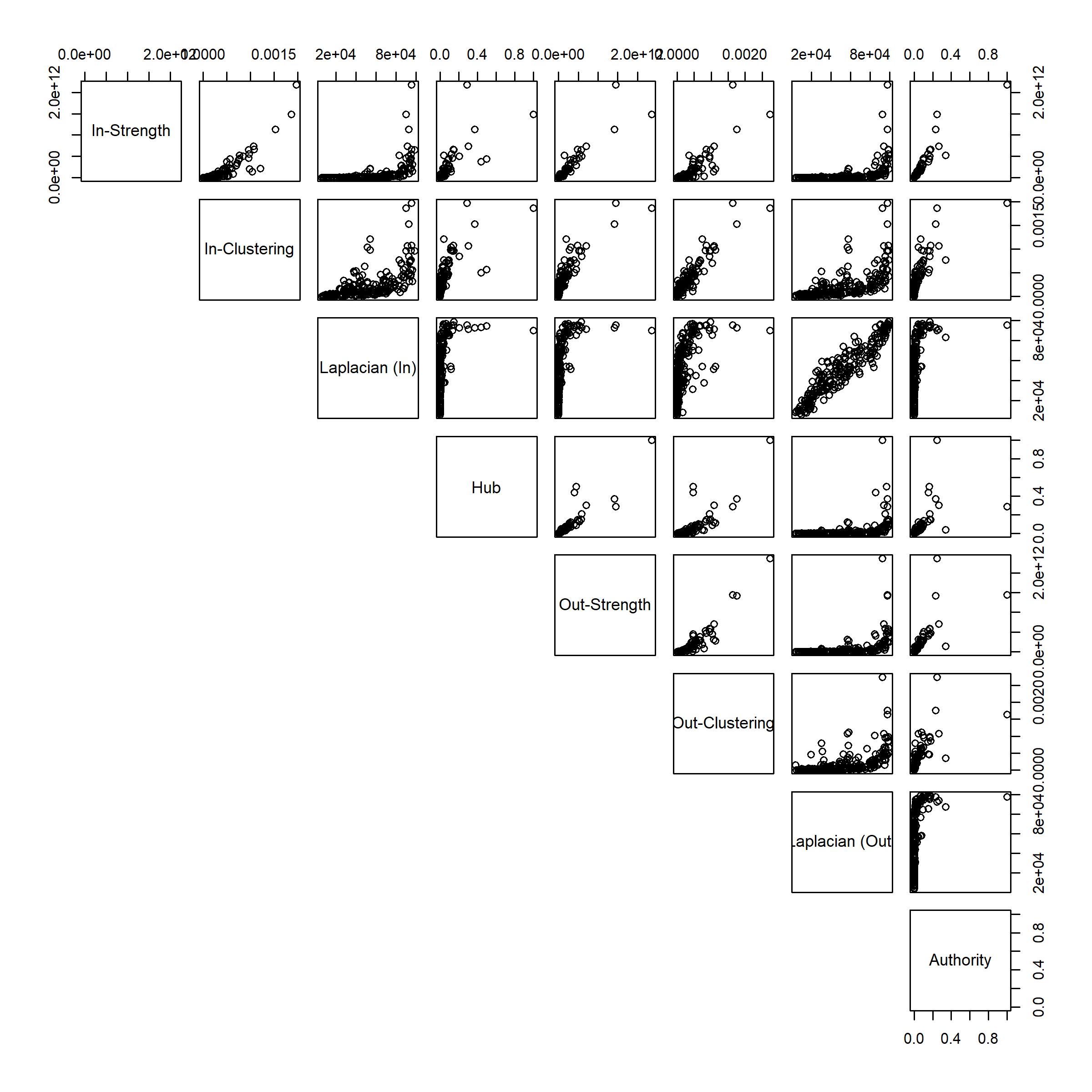}
	\caption{On the left-hand side, spearman correlation between each couple of measures. On the right-hand side, matrix of scatter plots between different indicators.}
	\label{fig:corr}
\end{figure}

As described in Section \ref{sec:model}, 
we aggregate the centrality indexes through a community detection method. As a result, communities are determined by the Clique Partition model, whose input is a weighted network constructed by  the original one,
in which weights are determined taking into account all the topological indicators in a multi-criteria approach. 
Four class of network indicators are initially computed by using the network depicted in Figure \ref{fig:network}. We report in Figure \ref{fig:corr} the scatter plots of each couple of centrality measures and the Spearman's rank-order correlation, in order to assess the strength and the direction of association between different ranked indicators. All the correlation are positive, because a country with a high volume of exports is also highly interconnected in the network. However, there are not  fully correlated couples and, in many cases, the correlation is far from one. It is also noteworthy the strong dependence between in and out versions of the same indicator. This is mainly explained by the similar patterns of imports and exports for several countries. Only hubs and authorities seem to emphasize the presence of specific exceptions. Table \ref{tab:table2} reports the top ten countries according to the rankings of the four used indicators. The rankings reflect the results about the correlations and they exemplify the differences in the role of each country as importer or exporter.

\begin{table}[!h]
	\centering
	\tiny
	\begin{tabular}{| c| c| c| c| c| c| c| c|}
		\hline		\hline		
		Laplacian In & Laplacian Out & In-Strength & Out-Strength & In-Clustering & Out-Clustering & Hubs & Authority\\
		\hline		\hline
		
		FRA	&	THA	&	USA	&	CHN	&	USA	&	CHN	&	CHN	&	USA	\\
		SGP	&	BLX	&	CHN	&	USA	&	CHN	&	DEU	&	CAN	&	HKG	\\	
		CZE	&	NLD	&	DEU	&	DEU	&	DEU	&	USA	&	MEX	&	JPN	\\
		USA	&	FRA	&	JPN	&	JPN	&	ARE	&	JPN	&	DEU	&	CHN	\\
		GBR	&	GBR	&	GBR	&	KOR	&	GBR	&	SAU	&	JPN	&	DEU	\\
		POL	&	DEU	&	FRA	&	FRA	&	JPN	&	RUS	&	USA	&	GBR	\\
		BLX	&	USA	&	NLD	&	NLD	&	SAU	&	FRA	&	KOR	&	KOR	\\
		NLD	&	SGP	&	HKG	&	ITA	&	NLD	&	ITA	&	FRA	&	FRA	\\
		THA	&	ITA	&	KOR	&	GBR	&	ITA	&	KOR	&	GBR	&	CAN	\\
		CAN	&	CAN	&	ITA	&	RUS	&	FRA	&	GBR	&	ITA	&	MEX	\\ \hline \hline
	\end{tabular}
	\caption{The top ten countries for each network indicator.}
	\label{tab:table2}	
\end{table}

\begin{figure}[!h]
	\centering
	\includegraphics[scale=0.4]{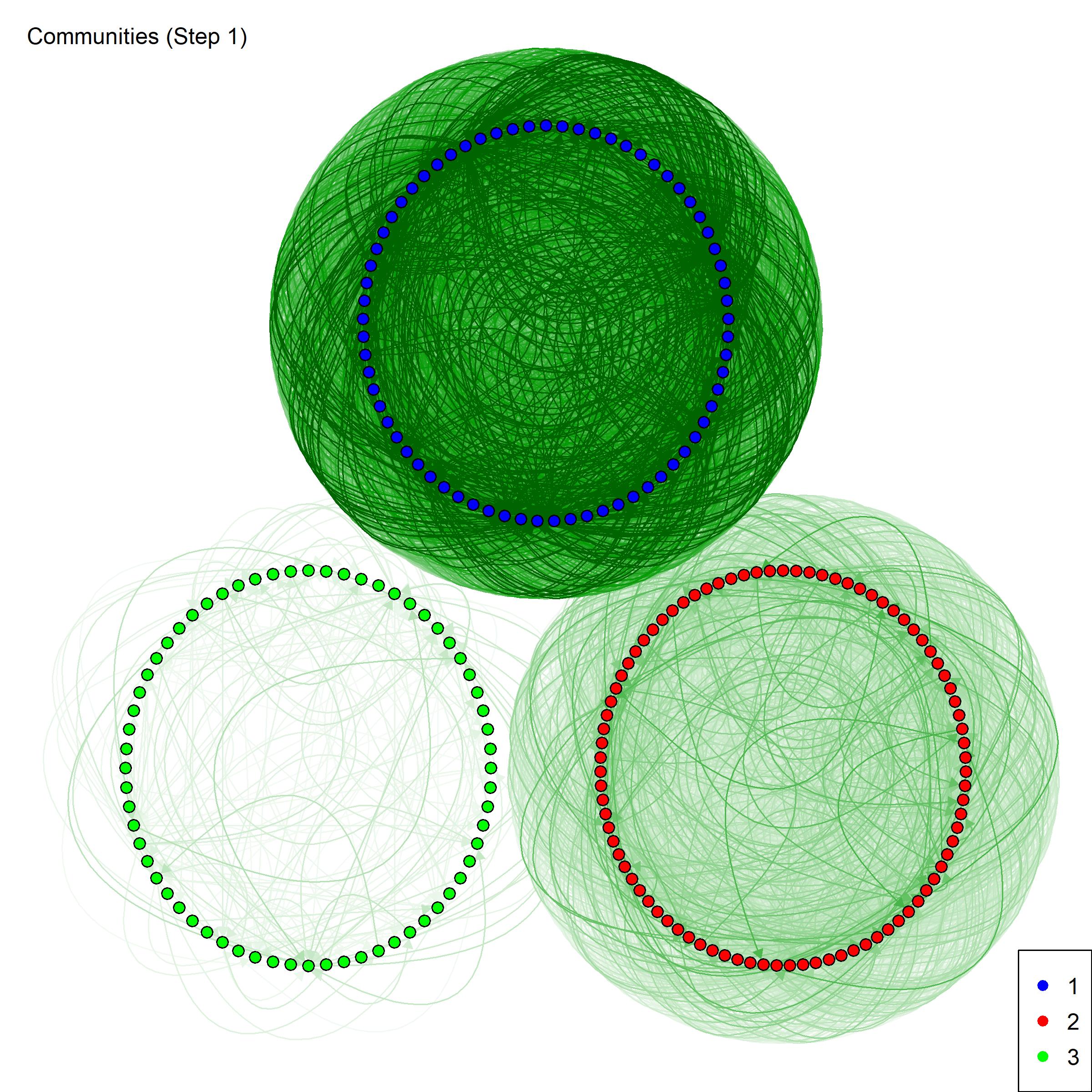}
	\caption{Clusters of countries identified at the first step by the community detection algorithm. The communities are ordered in terms of average ranking.}
	\label{fig:commstep1}
\end{figure}

By applying the methodology\footnote{In the application we set $p=2$ for the computation of the Minkoski distance. Similar results have been obtained by using other values of $p$.} described in Section \ref{sec:model}, we obtain at the first step three communities, characterized by 69, 87 and 64 countries, respectively. We display in Figures \ref{fig:commstep1}  the communities initially identified by the algorithm. These three clusters are also well separated in terms of countries' centrality. We have indeed that countries belonging to community 1 have an average ranking of 38, the second community has an average ranking of 113, while countries that belong to the lowest community have an average ranking around 185. In other words, the most central countries are all included in the top community.
We also notice that the three clusters are characterized by a very different intra-group density. We have indeed that the density of the subgraphs (of the original ITN) induced by the countries belonging to the three clusters is 0.97, 0.53, 0.05, respectively. This behaviour can be partially explained by the fact that central countries tend to concentrate a high number of transactions between them. \\

\begin{figure}
	\centering
	\includegraphics[scale=0.4]{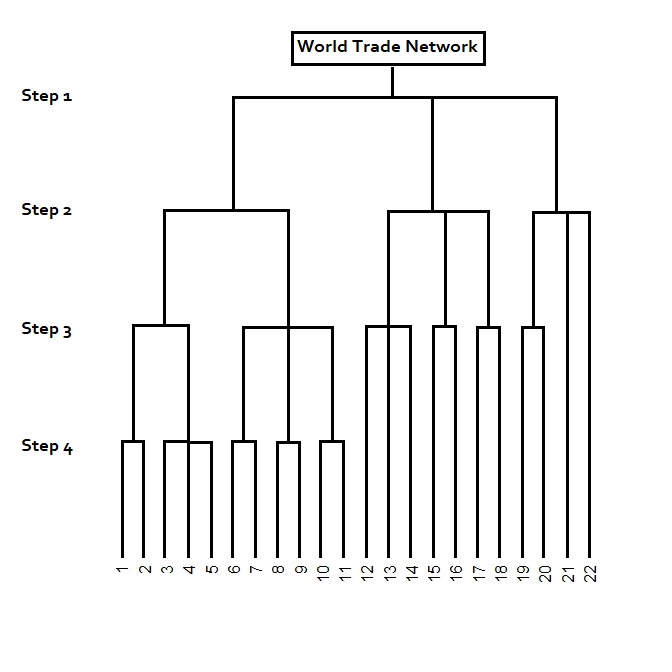}
	\caption{Dendrogram that illustrates the arrangement of clusters by applying the algorithm at four different levels. Communities are ordered in terms of average ranking.}
	\label{fig:dendrogram}
\end{figure}

\begin{figure}
	\centering
	\includegraphics[scale=0.25]{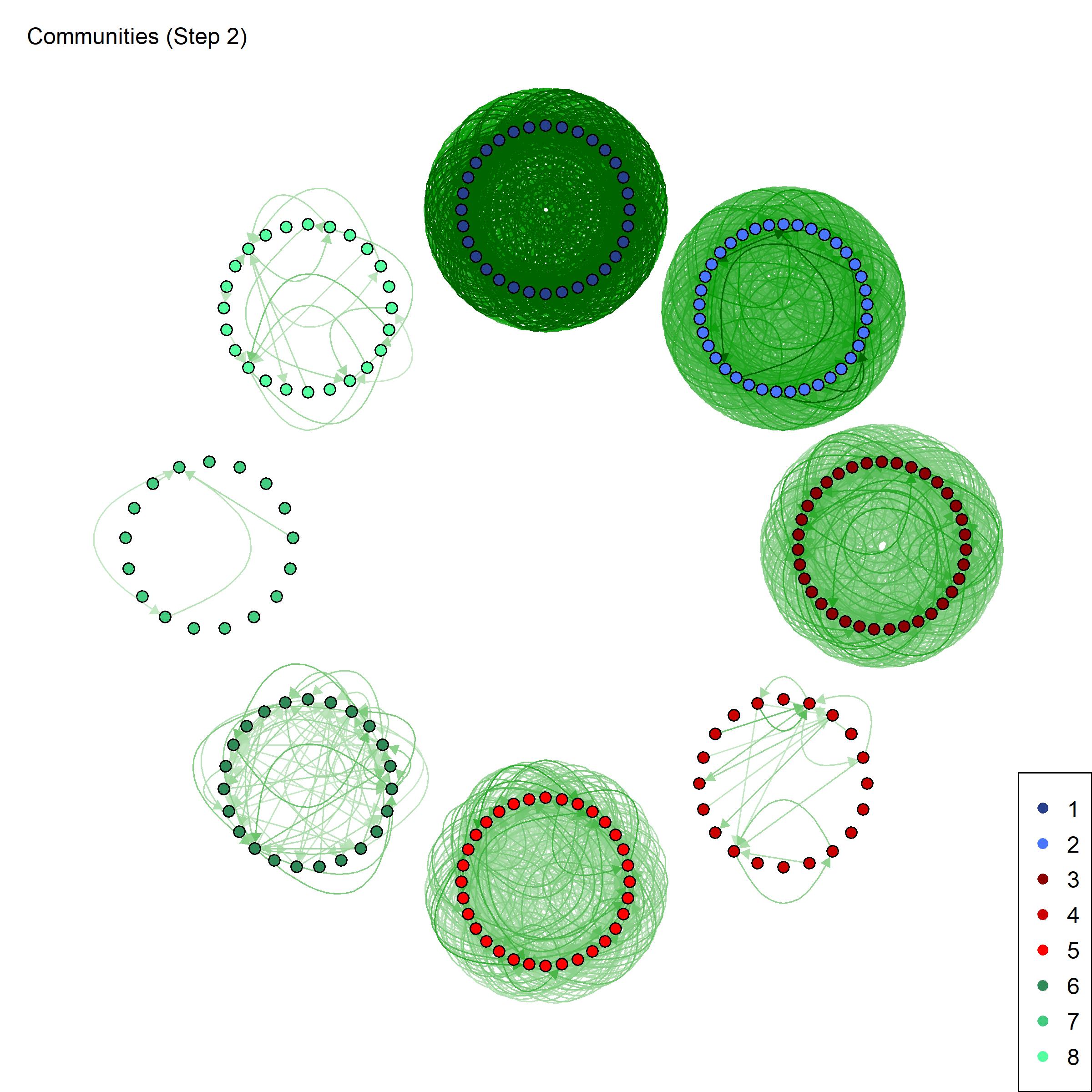}
	\includegraphics[scale=0.25]{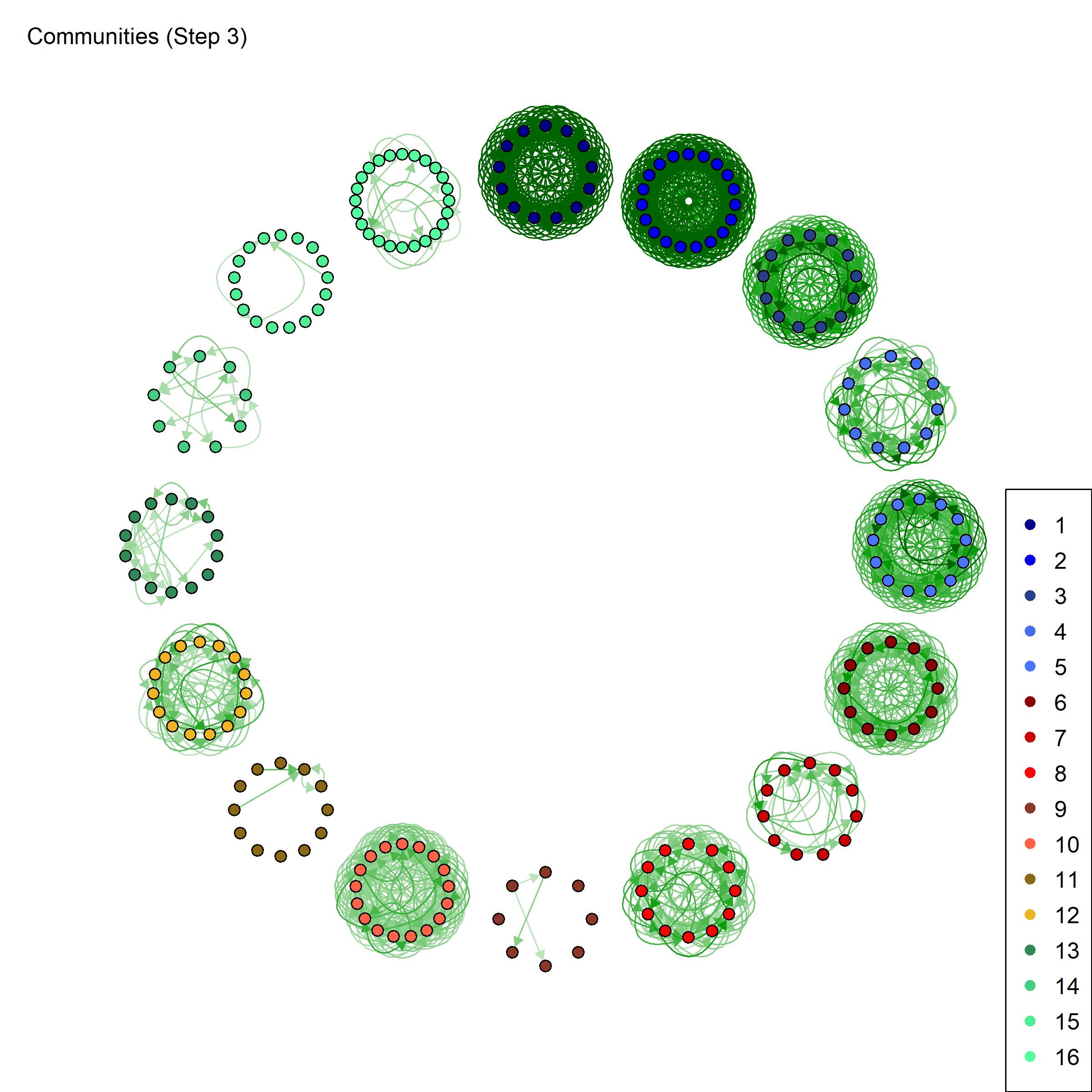}
	\includegraphics[scale=0.30]{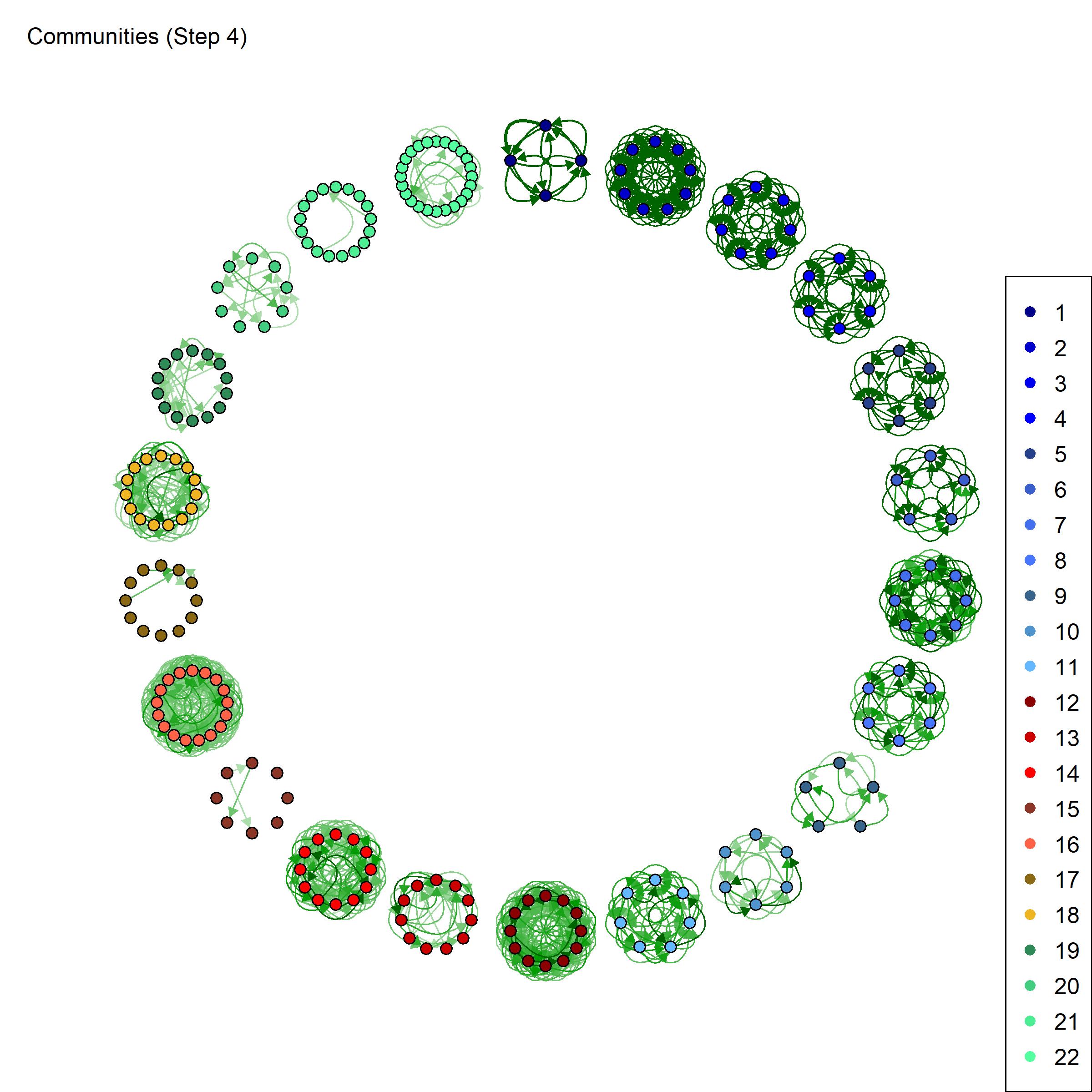}
	\caption{Clusters of countries identified at the second, third and fourth step, respectively, by the community detection algorithm. The communities are ordered in terms of average ranking.}
	\label{fig:communitystep2}
\end{figure}

Since in several contexts this initial division could be too raw, we can refine the procedure in order to reduce the heterogeneity in each group. To this end, at the subsequent step, we separately consider the ranking of centralities of countries, 
applying the proposed method for community detection to the single group. 
Specifically, at step 2  
we apply the proposed algorithm 
within each community detected at the previous step. In other words, at this step the algorithm takes into account how a specific country is ranked with respect to other countries of the same subgroup on the basis of the centrality indicators computed on the whole network. The ranking position of each country may change, but the global ranking remains the original one.
For instance, the community 1, characterized by 69 countries, splits into two groups of 32 and 37 countries, respectively. The two groups obtained have an average ranking of 19 and 55. The procedure is repeated in a similar way also for the other two communities identified at the step 1, resulting in 8 communities at step 2 (see dendrogram in Figure \ref{fig:dendrogram} and top left-hand side in Figure \ref{fig:communitystep2}). \\

Further reductions of the heterogeneity in each cluster are possible of course, repeating  again this process at the next steps and, in general, a stopping criterion is needed.
A possible one consists in looking at the volatility of the ranking inside each cluster. If we focus on community with larger standard deviation, we tend to produce a more refined breakdown between low-ranking countries. Vice versa, looking at a measure of relative volatility (as the coefficient of variation (CV)), we deal with a higher decomposition of top-ranking clusters. Here we follow this second approach and, at each step, we further divide a community only if the CV of countries' average rankings is higher than 7.5\%. \\
The complete structure representing the various division steps is represented by the dendrogram in Figure \ref{fig:dendrogram}. We notice that the number of communities increases at each step, leading to 22 communities at step 4. As expected, the criterion based on CV leads to a more granular breakdown for clusters characterized by a higher average ranking. In this way, we are able to classify key countries in different clusters. In Figure \ref{fig:communitystep2} we report the subnetworks induced by the clusters. The analysis confirms a tendency of top communities in showing a higher intra-group density. For instance, the top community at step 3 and the three higher ranking communities at step 4 are complete, that is all central countries trade each other. 
However, there is not a monotonic behaviour between ranking and intra-density. For instance, at step 2 community 4 has a higher average ranking than community 5 (124 against 128), but a significant lower intra density (0.05 against 0.58). This peculiar behaviour can be justified by the composition of the groups\footnote{Community 5 at step 2 is indeed characterized by various groups of countries that trade each other. For instance, in this group, we have several countries, originated after the breakup of Jugoslavia and Russia.}. Indeed, we are grouping countries on the basis of similarity in terms of their central role in the network instead of using preferential economic relationships.

\newpage
It is worth to compare our results with a well-known country-classification method based on the Economic Complexity Index (ECI). This index, introduced by \cite{Hausmann2014}, allows to rank countries in the ITN according to the diversification of their export flows, which reflects the amount of knowledge that drives their growth. The higher is the ECI, the more advanced and diversified is an economy. In particular, countries whose economic complexity is greater than expected (on the basis of their global income), tend to grow faster than rich countries with a low ECI. In this perspective, ECI represents a suitable tool for comparing countries in the ITN independently of their total output and it provides an independent measure of similarity. For instance, in Table \ref{tab:table}, we list the values of the ECI for the countries in the top four clusters detected. As shown in Table \ref{tab:tableMeanSD}, the mean value of such an index for each cluster is positively correlated with their ranking in the final partition we found at step 4. However, some exceptions are noticeable. For instance, China, in cluster 1, is characterised by a lower ECI than some countries in cluster 2 (e.g. UK and Italy) because of a lower diversification of exported commodities. Indeed, its wealth comes from a more homogeneous set of assets than UK and Italy, which can express a wider diversification in their total output. This could explain why the Standard Deviation inside each one of our communities is significantly high.

\begin{figure}[!h]
\centering
\includegraphics[scale=0.25]{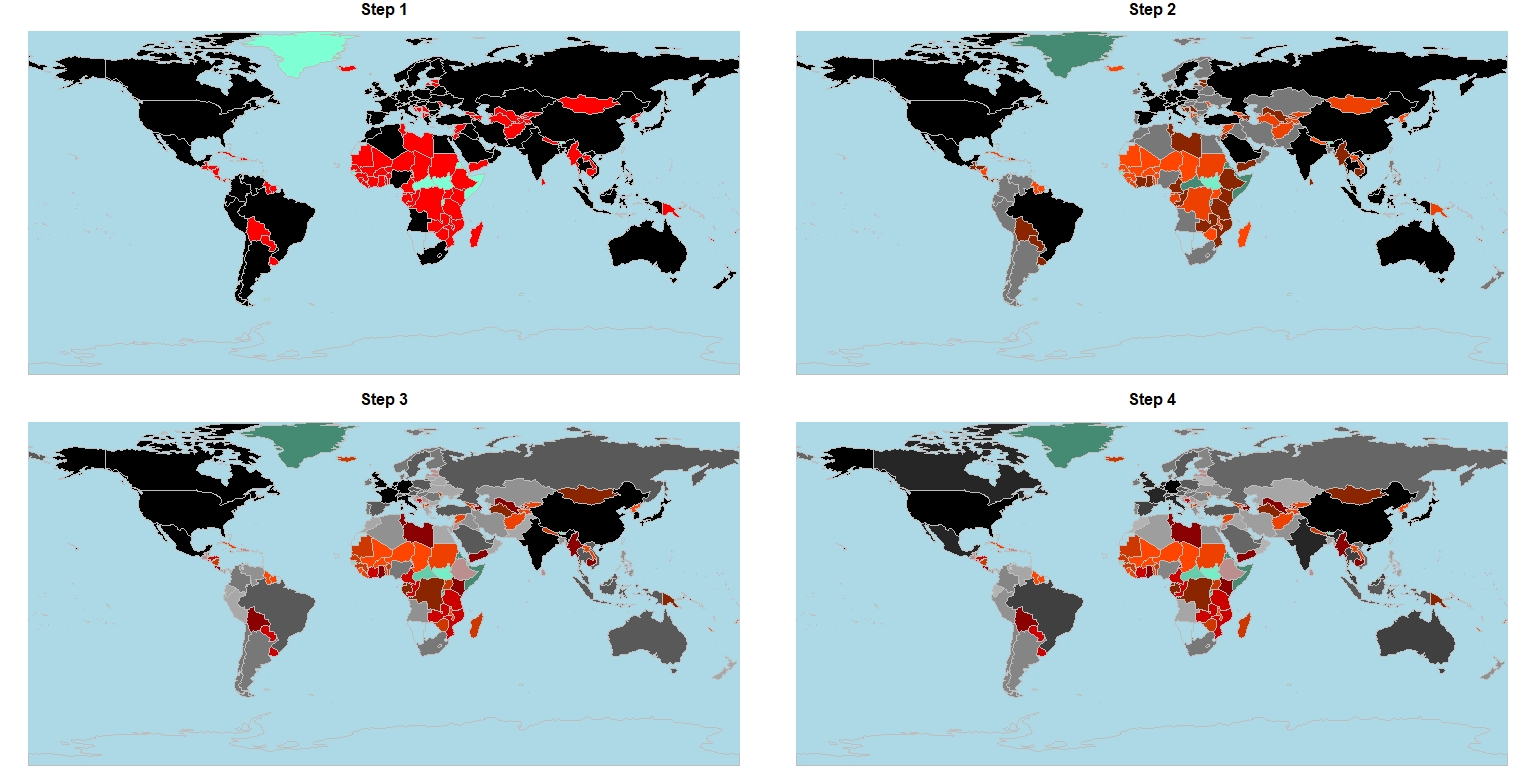}
\caption{Structure of communities at different steps. Darker colours are associated to communities with an higher average ranking. The number of communities is respectively equal to 3, 8, 16, 22.}
\label{fig:communityWorld}
\end{figure}

Now, we focus on the countries' role within the network. As shown in Figure \ref{fig:communityWorld}, the initial breakdown in communities gives a general feeling of the relevance of different macro-regions in the whole trade network. We have indeed that the top cluster, characterized by 69 countries at step 1, includes all the most developed European countries\footnote{28 European Countries are included in community 1. Gibraltar, San Marino and Andorra and some countries originated after the breakup of Jugoslavia and Russia are not included.}, largest economies in Asia and Middle East, several countries in South America, Canada, Mexico, USA, Australia and New Zealand. Furthermore, Algeria, Angola, Egypt, Morocco, Nigeria and South Africa are included for the African continent. Except for some small countries, this community includes all the advanced economies identified in the World Economic Outlook (WEO) by the International Monetary Fund (IMF)\footnote{List of advanced countries according to WEO are available at:
	
	https://www.imf.org/external/pubs/ft/weo/2019/01/weodata/groups.htm\#ea}
and the emerging economies identified by IMF and by other analysts\footnote{Various sources list countries as ``emerging economies'' exist. A few countries appear in every list (BRICS, Mexico, Turkey). While there are no commonly agreed upon parameters on which the countries can be classified as ``Emerging Economies'', several firms have developed detailed methodologies to identify the top performing emerging economies every year.}. \\
%Following steps produce a more granular division of countries. In particular in Table \ref{tab:table}, we report how community 1 is separated by the algorithm at different steps. 
At the end of the procedure, we obtain that the most central group is composed by China, Germany, Japan and United States. Higher volumes of trades are indeed moved by these countries (e.g., see ranking of in and out-strength in Table  \ref{tab:table2}) and, at the same time, they also show the highest levels of interconnections.  \\ In the second group, we have countries which either are positioned at a slightly lower level (as GBR, FRA, ITA and NLD) or are outstanding for one specific indicator, but, on average, they show a less relevant role in the network. For instance, Canada has the second position in terms of hubs centrality (see Table \ref{tab:table2}), but shows an average ranking around 14, because of a lower clustering. This is in line with its low value of the ECI. \\

\begin{table}[!h]
	\centering
	\tiny
	\begin{tabular}{|l| c| c| c| c|| c||}
		\hline		\hline		
		Country & Step 1 & Step 2 & Step 3 & Step 4 & ECI \\
		\hline		\hline
		CHN&	1&	1&	1&	1 & 1.16379 \\
		DEU&	1&	1&	1&	1 & 1.81367 \\
		JPN&	1&	1&	1&	1 & 2.31842 \\
		USA&	1&	1&	1&	1 & 1.30167 \\ \hdashline
		BLX&	1&	1&	1&	2 & 0.90581 \\
		CAN&	1&	1&	1&	2 & 0.411362\\
		FRA&	1&	1&	1&	2 & 1.15748 \\
		GBR&	1&	1&	1&	2 & 1.40296 \\
		IND&	1&	1&	1&	2 & -0.014696 \\
		ITA&	1&	1&	1&	2 & 1.24155 \\
		KOR&	1&	1&	1&	2 & 1.90646 \\
		MEX&	1&	1&	1&	2 & 0.953003 \\
		NLD&	1&	1&	1&	2 & 0.756212 \\ \hdashline
		AUS&	1&	1&	2&	3 & -0.846322 \\
		BRA&	1&	1&	2&	3 & -0.151225 \\
		CHE&	1&	1&	2&	3 & 1.99456 \\
		ESP&	1&	1&	2&	3 & 0.701443 \\
		MYS&	1&	1&	2&	3 & 0.828817 \\
		SGP&	1&	1&	2&	3 & 1.71171	\\
		THA&	1&	1&	2&	3 & 0.955651 \\ \hdashline
		AUT&	1&	1&	2&	4 & 1.64981 \\
		CZE&	1&	1&	2&	4 & 1.52129 \\
		IDN&	1&	1&	2&	4 & -0.014696 \\
		POL&	1&	1&	2&	4 & 0.839266 \\
		SWE&	1&	1&	2&	4 & 1.6459 \\
		TUR&	1&	1&	2&	4 & 0.378481 \\
		ARE&	1&	1&	2&	4 & -0.502072 \\
		HKG&	1&	1&	2&	4 & 1.35236 \\
		RUS&	1&	1&	2&	4 & 0.008439 \\
		SAU&	1&	1&	2&	4 & -0.369927 \\
		VNM&	1&	1&	2&	4 & -0.129961 \\
		XXB&	1&	1&	2&	4 & NA\\ \hline \hline
	\end{tabular}
	\caption{Composition of top four clusters (in terms of average ranking) derived at step 4. Last column displays the ECI for each country.}
	\label{tab:table}	
\end{table}

\begin{table}[!h]
	\centering
	\tiny
	\begin{tabular}{||c|c|c||}
		\hline		\hline		
		Community & Mean ECI & SD ECI\\
		\hline		\hline
		
		1	& 1.6493875  & 0.526404666 \\ \hline
		2	& 0.968904556 & 0.559587598 \\ \hline
		3	& 0.742090571 & 0.990344256 \\ \hline
		4	& 0.579899091 & 0.844314087 \\ \hline \hline
	\end{tabular}
	\caption{Mean and standard deviation of ECI inside each of the four top clusters}
	\label{tab:tableMeanSD}
\end{table}

\newpage
It is worth briefly comparing our results with those obtained by other community detection methods on the same network (see \cite{Barigozzi2011}, \cite{Piccardi2012} and \cite{BCG2020}). In particular, in \cite{Piccardi2012}, %several approaches have been proposed to analyse the community structure of the ITN at different times. B
	both directed and undirected networks have been tested without significant differences.  In \cite{BCG2020}, the authors follows an approach based on the maximisation of a specific quality function defined for general metric spaces. A quantitative correlation between the world partition in communities obtained by a modularity criterion and geographical distances has been investigated in \cite{Barigozzi2011}.  A common point of these alternative approaches is that the applied methodologies focus on the strength of countries' relationship in order to group togheter countries that trade each other. As a consequence, a common result is that geographical proximity still matters for international trade, jointly with trade agreements, common language or religion, and traditional partnerships. In all cases, a large relevant community including China and North America is observed. \\
As described in section \ref{nov}, the methodology proposed in this paper follows a different path for identifying clusters based on the relevance of countries in the network. Results display indeed in the same community countries that have an analogous role in the network. Hence, it could be interesting to compare them with papers that study how countries are positioned in the ITN. In this field, main approaches in the literature are based on the application of alternative centrality measures and main results show how different centrality measures, catching alternative aspects of the network structure, can provide a different ranking (see, e.g.,  \cite{Cingolani2017}). In this context, the main advantage of our approach is that we take jointly into account several indicators considering the peculiarities and the heterogeneity of different measures and we group togheter countries with a similar role according to the considered features. Comparing the results, we observe that the four countries (China, Germany, Japan and United States) that belong to the most central group, are on average also in the top positions of the economic sectors explored in \cite{Cingolani2017}. Similarly we found, in our second group, countries (as Mexico, Canada and South Korea) that  in \cite{Cingolani2017} appear to follow an intermediary role, having connections with both focal countries and less central ones. To conclude, it seems that the proposed approach is able to catch different elements of the network structure, providing, at the same time, a univocal classification of countries in terms of their relevance.

\section{Conclusions}\label{sec:conc}

Community detection is a widely discussed topic in network theory. The analysis of the mesoscale structure of a real network throws light on its inner structure. This plays an even more significant role when applied to ITN, in view of its multiple implications. 
This work aimed at clustering countries according to similarities in their role in the global market, rather than using only the preferential channels of exchange between them.
Centrality measures have represented, by now, a classical tool to rank such a role in the network. In particular, each centrality measure expresses a different information about the nodes position. 
We proposed a way to collect all the information content, represented by suitable centrality measures, through a distance measure between countries.

Among all possible similarity-dissimilarity distances, the Minkowski distance allows to grasp different data distributions, depending on a specific parameter $p$. 
In this way, we constructed a weighted complete network where nodes are countries and weighted links are related to similarities between them. By means of this similarity-network, we set up a classical Clique Partitioning problem to identify the community structure that maximizes the modularity.
We proposed here a new algorithm which, loosely speaking, merges different nodes or clusters and shrinks the network in such a way to get polynomial times for its solution.

When applied to the ITN in the year 2014, the optimal solution shows three big clusters, more or less equivalent in size but very different in terms of intra-cluster density. This has been easily interpreted since the rate of exchanges between top countries is far more intense than for poor ones. We iterated the same methodology to each cluster, in order to reduce the internal heterogeneity.
This allows to build a dendrogram tree stemming at each step. 

The top leader economies in the world result to be those of China, Japan, USA and Germany. This is not unexpected but our proposal shows that these countries also play a very similar role in the world economy on the basis of the set of selected indicators, making our approach suitable for other network applications.

%\section{Section title}
%\label{sec:1}
%Text with citations \cite{RefB} and \cite{RefJ}.
%\subsection{Subsection title}
%\label{sec:2}
%as required. Don't forget to give each section
%and subsection a unique label (see Sect.~\ref{sec:1}).
%\paragraph{Paragraph headings} Use paragraph headings as needed.
%\begin{equation}
%\end{equation}

%\begin{acknowledgements}
%\end{acknowledgements}


 \section*{Conflict of interest}
 The authors declare that they have no conflict of interest.

\bibliography{Myref}   % name your BibTeX data base

\begin{thebibliography}{70}
\providecommand{\natexlab}[1]{#1}
\providecommand{\url}[1]{\texttt{#1}}
\expandafter\ifx\csname urlstyle\endcsname\relax
  \providecommand{\doi}[1]{doi: #1}\else
  \providecommand{\doi}{doi: \begingroup \urlstyle{rm}\Url}\fi

\bibitem[Adiga and Smitha(2009)]{Adiga}
C.~Adiga and M.~Smitha.
\newblock {On the skew Laplacian energy of a digraph}.
\newblock \emph{International Mathematical Forum}, 4\penalty0 (3):\penalty0
  1907--1914, 2009.

\bibitem[Aloise et~al.(2010)Aloise, Cafieri, Caporossi, Hansen, Perron, and
  Liberti]{Aloise2010}
D.~Aloise, S.~Cafieri, G.~Caporossi, P.~Hansen, S.~Perron, and L.~Liberti.
\newblock Column generation algorithms for exact modularity maximization in
  networks.
\newblock \emph{Physical Review E}, 82\penalty0 (4):\penalty0 046112, 2010.

\bibitem[Barbero and Zof{\'\i}o(2016)]{Barbero2016}
J.~Barbero and J.~L. Zof{\'\i}o.
\newblock The multiregional core-periphery model: The role of the spatial
  topology.
\newblock \emph{Networks and Spatial Economics}, 16\penalty0 (2):\penalty0
  469--496, 2016.

\bibitem[Barigozzi et~al.(2011)Barigozzi, Fagiolo, and Mangioni]{Barigozzi2011}
M.~Barigozzi, G.~Fagiolo, and G.~Mangioni.
\newblock Identifying the community structure of the international-trade
  multi-network.
\newblock \emph{Physica A: statistical mechanics and its applications},
  390\penalty0 (11):\penalty0 2051--2066, 2011.

\bibitem[Barrat et~al.(2004)Barrat, Barth\'elemy, Pastor-Satorras, and
  Vespignani]{Barrat_2004}
A.~Barrat, M.~Barth\'elemy, R.~Pastor-Satorras, and A.~Vespignani.
\newblock The architecture of complex weighted networks.
\newblock \emph{Proceedings of the National Academy of Sciences}, 101\penalty0
  (11):\penalty0 3747--3752, 2004.

\bibitem[Bartesaghi et~al.(2020)Bartesaghi, Clemente, and Grassi]{BCG2020}
P.~Bartesaghi, G.~P. Clemente, and R.~Grassi.
\newblock Community structure in the world trade network based on
  communicability distances.
\newblock \emph{Journal of Economic Interaction and Coordination}, 2020.

\bibitem[Baruah and Bharali(2017)]{Baruah}
D.~Baruah and A.~Bharali.
\newblock {A Comparative Study of Vertex Deleted Centrality Measures}.
\newblock \emph{Annals of Pure and Applied Mathematics}, 14\penalty0
  (1):\penalty0 199--205, 2017.

\bibitem[Bl{\"o}chl et~al.(2011)Bl{\"o}chl, Theis, Vega-Redondo, and
  Fisher]{Blochl2011}
F.~Bl{\"o}chl, F.~J. Theis, F.~Vega-Redondo, and E.~O. Fisher.
\newblock Vertex centralities in input-output networks reveal the structure of
  modern economies.
\newblock \emph{Physical Review E}, 83\penalty0 (4):\penalty0 046127, 2011.

\bibitem[Blondel et~al.(2008)Blondel, Guillaume, Lambiotte, and
  Lefebvre]{blondel2008fast}
V.~D. Blondel, J.-L. Guillaume, R.~Lambiotte, and E.~Lefebvre.
\newblock Fast unfolding of communities in large networks.
\newblock \emph{Journal of statistical mechanics: theory and experiment},
  2008\penalty0 (10):\penalty0 P10008, 2008.

\bibitem[Brandes and Erlebach(2005)]{Brandes}
U.~Brandes and T.~Erlebach.
\newblock \emph{Network Analysis. Methodological Foundations}.
\newblock Springer, 2005.

\bibitem[Brusco and K{\"o}hn(2009)]{Brusco09}
M.~J. Brusco and H.-F. K{\"o}hn.
\newblock Clustering qualitative data based on binary equivalence relations:
  neighborhood search heuristics for the clique partitioning problem.
\newblock \emph{Psychometrika}, 74\penalty0 (4):\penalty0 685, 2009.

\bibitem[Butenko and Wilhelm(2006)]{BUTENKO2005}
S.~Butenko and W.~Wilhelm.
\newblock Clique-detection models in computational biochemistry and genomics.
\newblock \emph{European Journal of Operational Research}, 173:\penalty0 1 --
  17, 2006.

\bibitem[Cepeda-L{\'o}pez et~al.(2019)Cepeda-L{\'o}pez, Gamboa-Estrada,
  Le{\'o}n, and Rinc{\'o}n-Castro]{cepeda}
F.~Cepeda-L{\'o}pez, F.~Gamboa-Estrada, C.~Le{\'o}n, and H.~Rinc{\'o}n-Castro.
\newblock The evolution of world trade from 1995 to 2014: A network approach.
\newblock \emph{The Journal of International Trade \& Economic Development},
  28\penalty0 (4):\penalty0 452--485, 2019.

\bibitem[Cerqueti et~al.(2018)Cerqueti, Ferraro, and Iovanella]{Roy}
R.~Cerqueti, G.~Ferraro, and A.~Iovanella.
\newblock A new measure for community structure through indirect social
  connections.
\newblock \emph{Expert Systems with Applications}, 114:\penalty0 196--209,
  2018.

\bibitem[Chelouah and Siarry(2000)]{CHELOUAH2000}
R.~Chelouah and P.~Siarry.
\newblock Tabu search applied to global optimization.
\newblock \emph{European Journal of Operational Research}, 123:\penalty0 256 --
  270, 2000.

\bibitem[Cingolani et~al.(2017)Cingolani, Panzarasa, and Tajoli]{Cingolani2017}
I.~Cingolani, P.~Panzarasa, and L.~Tajoli.
\newblock Countries’ positions in the international global value networks:
  Centrality and economic performance.
\newblock \emph{Applied network science}, 2\penalty0 (1):\penalty0 21, 2017.

\bibitem[Clauset et~al.(2004)Clauset, Newman, and Moore]{clauset2004finding}
A.~Clauset, M.~E. Newman, and C.~Moore.
\newblock Finding community structure in very large networks.
\newblock \emph{Physical review E}, 70\penalty0 (6):\penalty0 066111, 2004.

\bibitem[Clauset et~al.(2008)Clauset, Moore, and
  Newman]{Clauset2008hierarchical}
A.~Clauset, C.~Moore, and M.~E. Newman.
\newblock Hierarchical structure and the prediction of missing links in
  networks.
\newblock \emph{Nature}, 453\penalty0 (7191):\penalty0 98--101, 2008.

\bibitem[Clemente and Grassi(2018)]{CleGra}
G.~P. Clemente and R.~Grassi.
\newblock {Directed clustering in weighted networks: A new perspective}.
\newblock \emph{Chaos, Solitons \& Fractals}, 107:\penalty0 26--38, 2018.

\bibitem[Danon et~al.(2006)Danon, D{\'\i}az-Guilera, and Arenas]{Danon2006}
L.~Danon, A.~D{\'\i}az-Guilera, and A.~Arenas.
\newblock The effect of size heterogeneity on community identification in
  complex networks.
\newblock \emph{Journal of Statistical Mechanics: Theory and Experiment},
  2006\penalty0 (11):\penalty0 P11010, 2006.

\bibitem[de~Amorim and Mirkin(2012)]{DeAmorim2012}
R.~C. de~Amorim and B.~Mirkin.
\newblock Minkowski metric, feature weighting and anomalous cluster
  initializing in k-means clustering.
\newblock \emph{Pattern Recognition}, 45:\penalty0 1061--1075, 2012.

\bibitem[de~Amorim et~al.(1992)de~Amorim, Barth\'{e}lemy, and
  Ribeiro]{deAmorim1992}
S.~de~Amorim, J.-P. Barth\'{e}lemy, and C.~Ribeiro.
\newblock Clustering and clique partitioning: Simulated annealing and tabu
  search approaches.
\newblock \emph{Journal of Classification}, 9\penalty0 (1):\penalty0 17--41,
  1992.

\bibitem[De~Benedictis and Tajoli(2011)]{DeBenedictis2011}
L.~De~Benedictis and L.~Tajoli.
\newblock The world trade network.
\newblock \emph{The World Economy}, 34\penalty0 (8):\penalty0 1417--1454, 2011.

\bibitem[De~Benedictis and Tajoli(2016)]{deb}
L.~De~Benedictis and L.~Tajoli.
\newblock {Comparative Advantage and Centrality in the World Network of Trade
  and Value Added: An Analysis of the Italian Position}.
\newblock \emph{Rivista di Politica Economica}, 66\penalty0 ((3)), 2016.

\bibitem[Fagiolo(2007)]{Fagiolo_2007}
G.~Fagiolo.
\newblock Clustering in complex directed networks.
\newblock \emph{Physical Review E}, 76\penalty0 (2), 2007.

\bibitem[Fagiolo et~al.(2008)Fagiolo, Reyes, and Schiavo]{FagioloR2008}
G.~Fagiolo, J.~Reyes, and S.~Schiavo.
\newblock On the topological properties of the world trade web: A weighted
  network analysis.
\newblock \emph{Physica A: Statistical Mechanics and its Applications},
  387\penalty0 (15):\penalty0 3868--3873, 2008.

\bibitem[Fagiolo et~al.(2010)Fagiolo, Reyes, and Schiavo]{Fagiolo2010}
G.~Fagiolo, J.~Reyes, and S.~Schiavo.
\newblock The evolution of the world trade web: a weighted-network analysis.
\newblock \emph{Journal of Evolutionary Economics}, 20\penalty0 (4):\penalty0
  479--514, 2010.

\bibitem[Ferraz~de Arruda et~al.(2014)Ferraz~de Arruda, Luiz~Barbieri,
  Rodríguez, Rodrigues, Moreno, and da~Fontoura~Costa]{Ferraz}
G.~Ferraz~de Arruda, A.~Luiz~Barbieri, P.~M. Rodríguez, F.~A. Rodrigues,
  Y.~Moreno, and L.~da~Fontoura~Costa.
\newblock The role of centrality for the identification of influential
  spreaders in complex networks.
\newblock \emph{Physical Review E}, 90, 2014.

\bibitem[Fortunato(2010)]{Fortunato2010}
S.~Fortunato.
\newblock Community detection in graphs.
\newblock \emph{Physics reports}, 486\penalty0 (3-5):\penalty0 75--174, 2010.

\bibitem[Fortunato and Hric(2016)]{FortunatoHric2016}
S.~Fortunato and D.~Hric.
\newblock Community detection in networks: A user guide.
\newblock \emph{Physics reports}, 659:\penalty0 1--44, 2016.

\bibitem[Garlaschelli and Loffredo(2004)]{Garlaschelli2004}
D.~Garlaschelli and M.~I. Loffredo.
\newblock Fitness-dependent topological properties of the world trade web.
\newblock \emph{Physical review letters}, 93\penalty0 (18):\penalty0 188701,
  2004.

\bibitem[Garlaschelli and Loffredo(2005)]{Garlaschelli2005}
D.~Garlaschelli and M.~I. Loffredo.
\newblock Structure and evolution of the world trade network.
\newblock \emph{Physica A: Statistical Mechanics and its Applications},
  355\penalty0 (1):\penalty0 138--144, 2005.

\bibitem[Garlaschelli et~al.(2007)Garlaschelli, Di~Matteo, Aste, Caldarelli,
  and Loffredo]{Garlaschelli2007}
D.~Garlaschelli, T.~Di~Matteo, T.~Aste, G.~Caldarelli, and M.~I. Loffredo.
\newblock Interplay between topology and dynamics in the world trade web.
\newblock \emph{The European Physical Journal B}, 57\penalty0 (2):\penalty0
  159--164, 2007.

\bibitem[Gaulier(2010)]{CEPII}
S.~Gaulier, G.;~Zignago.
\newblock {BACI: International Trade Database at the Product-Level. The
  1994-2007 Version}.
\newblock Technical Report 2010-23, CEPII,, 2010.

\bibitem[Gr\"{o}tschel and Wakabayashi(1989)]{Grotschel1989}
M.~Gr\"{o}tschel and Y.~Wakabayashi.
\newblock A cutting plane algorithm for a clustering problem.
\newblock \emph{Mathematical Programming}, 45\penalty0 (1-3):\penalty0 59--96,
  1989.

\bibitem[Gr\"{o}tschel and Wakabayashi(1990)]{Grotschel1990}
M.~Gr\"{o}tschel and Y.~Wakabayashi.
\newblock Facets of the clique partitioning polytope.
\newblock \emph{Mathematical Programming}, 47\penalty0 (1-3):\penalty0
  367--387, 1990.

\bibitem[Gutman and Zhou(2006)]{Gutman2006}
I.~Gutman and B.~Zhou.
\newblock Laplacian energy of a graph.
\newblock \emph{Linear Algebra and its applications}, 414\penalty0
  (1):\penalty0 29--37, 2006.

\bibitem[Haemers(1995)]{Haemers}
W.~Haemers.
\newblock {Interlacing eigenvalues and graphs}.
\newblock \emph{Linear Algebra and its Applications,}, 226-228:\penalty0
  593--616, 1995.

\bibitem[Hajdu et~al.(2019)Hajdu, B{\'o}ta, Kr{\'e}sz, Khani, and
  Gardner]{Hajdu2019}
L.~Hajdu, A.~B{\'o}ta, M.~Kr{\'e}sz, A.~Khani, and L.~M. Gardner.
\newblock Discovering the hidden community structure of public transportation
  networks.
\newblock \emph{Networks and Spatial Economics}, pages 1--23, 2019.

\bibitem[Hausmann et~al.(2014)Hausmann, Hidalgo, Bustos, Coscia, Simoes, and
  Yildirim]{Hausmann2014}
R.~Hausmann, C.~A. Hidalgo, S.~Bustos, M.~Coscia, A.~Simoes, and M.~A.
  Yildirim.
\newblock \emph{The atlas of economic complexity: Mapping paths to prosperity}.
\newblock Mit Press, 2014.

\bibitem[Kali and Reyes(2007)]{Kali2007}
R.~Kali and J.~Reyes.
\newblock The architecture of globalization: a network approach to
  international economic integration.
\newblock \emph{Journal of International Business Studies}, 38\penalty0
  (4):\penalty0 595--620, 2007.

\bibitem[Kali and Reyes(2010)]{Kali2010}
R.~Kali and J.~Reyes.
\newblock Financial contagion on the international trade network.
\newblock \emph{Economic Inquiry}, 48\penalty0 (4):\penalty0 1072--1101, 2010.

\bibitem[Kim and Shin(2002)]{Kim2002}
S.~Kim and E.-H. Shin.
\newblock A longitudinal analysis of globalization and regionalization in
  international trade: A social network approach.
\newblock \emph{Social forces}, 81\penalty0 (2):\penalty0 445--468, 2002.

\bibitem[Kissani and Mizoguchi(2010)]{Kissani}
P.~Kissani and Y.~Mizoguchi.
\newblock {Laplacian energy of directed graphs and minimizing maximum outdegree
  algorithms}.
\newblock Technical report, Kyushu University Institutional Repository, 2010.

\bibitem[Kleinberg(1999)]{Kleinberg1999}
J.~M. Kleinberg.
\newblock Authoritative sources in a hyperlinked environment.
\newblock \emph{Journal of the ACM (JACM)}, 46\penalty0 (5):\penalty0 604--632,
  1999.

\bibitem[Lazi{\'{c}}(2006)]{Lazic2006}
M.~Lazi{\'{c}}.
\newblock On the laplacian energy of a graph.
\newblock \emph{Czechoslovak Mathematical Journal}, 56\penalty0 (4):\penalty0
  1207--1213, Dec 2006.
\newblock ISSN 1572-9141.

\bibitem[Li et~al.(2003)Li, Jin, and Chen]{Li2003}
X.~Li, Y.~Y. Jin, and G.~Chen.
\newblock Complexity and synchronization of the world trade web.
\newblock \emph{Physica A: Statistical Mechanics and its Applications},
  328\penalty0 (1-2):\penalty0 287--296, 2003.

\bibitem[Mehrotra and Trick(1998)]{Mehrotra1998}
A.~Mehrotra and M.~Trick.
\newblock Cliques and clustering: A combinatorial approach.
\newblock \emph{Operations Research Letters}, 22\penalty0 (1):\penalty0 1--12,
  1998.

\bibitem[Newman(2004)]{newman2004Fast}
M.~E. Newman.
\newblock Fast algorithm for detecting community structure in networks.
\newblock \emph{Physical review E}, 69\penalty0 (6):\penalty0 066133, 2004.

\bibitem[Newman and Girvan(2004)]{Newman2004}
M.~E. Newman and M.~Girvan.
\newblock Finding and evaluating community structure in networks.
\newblock \emph{Physical review E}, 69\penalty0 (2):\penalty0 026113, 2004.

\bibitem[Newman(2010)]{Newman2010}
M.~E.~J. Newman.
\newblock \emph{Networks: an introduction}.
\newblock Oxford university press, 2010.

\bibitem[Onnela et~al.(2005)Onnela, Saram\"{a}ki, Kert{\'{e}}sz, and
  Kaski]{Onnela_2005}
J.~Onnela, J.~Saram\"{a}ki, J.~Kert{\'{e}}sz, and K.~Kaski.
\newblock Intensity and coherence of motifs in weighted complex networks.
\newblock \emph{Physical Review E}, 71\penalty0 (6), 2005.

\bibitem[Pattillo et~al.(2013)Pattillo, Youssef, and Butenko]{PATTILLO20139}
J.~Pattillo, N.~Youssef, and S.~Butenko.
\newblock On clique relaxation models in network analysis.
\newblock \emph{European Journal of Operational Research}, 226\penalty0
  (1):\penalty0 9 -- 18, 2013.

\bibitem[Piccardi(2011)]{Piccardi2011}
C.~Piccardi.
\newblock Finding and testing network communities by lumped markov chains.
\newblock \emph{PLOS ONE}, 6\penalty0 (11):\penalty0 1--13, 11 2011.

\bibitem[Piccardi and Tajoli(2012)]{Piccardi2012}
C.~Piccardi and L.~Tajoli.
\newblock Existence and significance of communities in the world trade web.
\newblock \emph{Phys. Rev. E}, 85:\penalty0 066119, Jun 2012.

\bibitem[Qi et~al.(2012)Qi, Fuller, Wu, Wu, and Zhang]{Qi2012}
X.~Qi, E.~Fuller, Q.~Wu, Y.~Wu, and C.-Q. Zhang.
\newblock Laplacian centrality: A new centrality measure for weighted networks.
\newblock \emph{Information Sciences}, 194:\penalty0 240--253, 2012.

\bibitem[Rotundo and Ausloos(2010)]{Rotundo2010}
G.~Rotundo and M.~Ausloos.
\newblock Organization of networks with tagged nodes and biased links: A priori
  distinct communities: The case of intelligent design proponents and
  {Darwinian} evolution defenders.
\newblock \emph{Physica A: Statistical Mechanics and its Applications},
  389\penalty0 (23):\penalty0 5479--5494, 2010.

\bibitem[Rudin(2009)]{Rudin2009}
C.~Rudin.
\newblock The p-norm push: A simple convex ranking algorithm that concentrates
  at the top of the list.
\newblock \emph{Journal of Machine Learning Research}, 10\penalty0
  (Oct):\penalty0 2233--2271, 2009.

\bibitem[Santiago and Lamb(2017)]{SANTIAGO2017844}
R.~Santiago and L.~C. Lamb.
\newblock Efficient modularity density heuristics for large graphs.
\newblock \emph{European Journal of Operational Research}, 258\penalty0
  (3):\penalty0 844 -- 865, 2017.
\newblock ISSN 0377-2217.

\bibitem[Schiavo et~al.(2010)Schiavo, Reyes, and Fagiolo]{Schiavo2010}
S.~Schiavo, J.~Reyes, and G.~Fagiolo.
\newblock International trade and financial integration: a weighted network
  analysis.
\newblock \emph{Quantitative Finance}, 10\penalty0 (4):\penalty0 389--399,
  2010.

\bibitem[Serrano and Bogu{\~n}{\'a}(2003)]{Serrano2003}
M.~A. Serrano and M.~Bogu{\~n}{\'a}.
\newblock Topology of the world trade web.
\newblock \emph{Physical Review E}, 68\penalty0 (1):\penalty0 015101, 2003.

\bibitem[Serrano et~al.(2007)Serrano, Bogu{\~n}{\'a}, and
  Vespignani]{Serrano2007}
M.~A. Serrano, M.~Bogu{\~n}{\'a}, and A.~Vespignani.
\newblock Patterns of dominant flows in the world trade web.
\newblock \emph{Journal of Economic Interaction and Coordination}, 2\penalty0
  (2):\penalty0 111--124, 2007.

\bibitem[Smith and White(1992)]{Smith1992}
D.~A. Smith and D.~R. White.
\newblock Structure and dynamics of the global economy: network analysis of
  international trade 1965--1980.
\newblock \emph{Social forces}, 70\penalty0 (4):\penalty0 857--893, 1992.

\bibitem[Snyder and Kick(1979)]{Snyder1979}
D.~Snyder and E.~L. Kick.
\newblock Structural position in the world system and economic growth,
  1955-1970: A multiple-network analysis of transnational interactions.
\newblock \emph{American journal of Sociology}, 84\penalty0 (5):\penalty0
  1096--1126, 1979.

\bibitem[Tzekina et~al.(2008)Tzekina, Danthi, and Rockmore]{tzekina2008}
I.~Tzekina, K.~Danthi, and D.~N. Rockmore.
\newblock Evolution of community structure in the world trade web.
\newblock \emph{The European Physical Journal B}, 63\penalty0 (4):\penalty0
  541--545, 2008.

\bibitem[Varela et~al.(2015)Varela, Rotundo, Ausloos, and Carrete]{Varela2015}
L.~M. Varela, G.~Rotundo, M.~Ausloos, and J.~Carrete.
\newblock Complex network analysis in socioeconomic models.
\newblock In \emph{Complexity and Geographical Economics}, pages 209--245.
  Springer, 2015.

\bibitem[Wang et~al.(2008)Wang, Obremski, Alidaee, and Kochenberger]{Wang2008}
H.~Wang, T.~Obremski, B.~Alidaee, and G.~Kochenberger.
\newblock Clique partitioning for clustering: A comparison with k-means and
  latent class analysis.
\newblock \emph{Communications in Statistics: Simulation and Computation},
  37\penalty0 (1):\penalty0 1--13, 2008.

\bibitem[Wasserman and Faust(1994)]{WasFaust}
S.~Wasserman and K.~Faust.
\newblock \emph{Social Network Analysis: Methods and Applications.}
\newblock Cambridge University Press, New York, NY., July 1994.

\bibitem[Watts and Strogatz(1998)]{Watts_1998}
D.~J. Watts and S.~H. Strogatz.
\newblock Collective dynamics of \lq small-world \rq networks.
\newblock \emph{Nature}, 393\penalty0 (6684):\penalty0 440--442, jun 1998.

\bibitem[Zhang et~al.(2017)Zhang, Cui, Zhu, Du, Wang, and Shi]{Zhang2017}
X.~Zhang, H.~Cui, J.~Zhu, Y.~Du, Q.~Wang, and W.~Shi.
\newblock Measuring the dissimilarity of multiplex networks: An empirical study
  of international trade networks.
\newblock \emph{Physica A: Statistical Mechanics and its Applications},
  467:\penalty0 380--394, 2017.

\end{thebibliography}

%\begin{thebibliography}{}
% etc
%\end{thebibliography}

\end{document}